\documentclass{jfm}
\pdfoutput=1
\usepackage{graphicx}
\usepackage{natbib}

\usepackage{amsmath}
\usepackage{amsbsy}
\usepackage{amssymb}
\usepackage{palatino,avant,graphicx,color}
\usepackage[none]{hyphenat}
\usepackage{tikz}
\usepackage{xy}
\usepackage{comment}
\usepackage[none]{changes}
\usepackage{wrapfig}
\usepackage{xpatch}

\ifCUPmtlplainloaded \else
  \checkfont{eurm10}
  \iffontfound
    \IfFileExists{upmath.sty}
      {\typeout{^^JFound AMS Euler Roman fonts on the system,
                   using the 'upmath' package.^^J}%
       \usepackage{upmath}}
      {\typeout{^^JFound AMS Euler Roman fonts on the system, but you
                   dont seem to have the}%
       \typeout{'upmath' package installed. JFM.cls can take advantage
                 of these fonts,^^Jif you use 'upmath' package.^^J}%
      }
  \else
  \fi
\fi

\ifCUPmtlplainloaded \else
  \checkfont{msam10}
  \iffontfound
    \IfFileExists{amssymb.sty}
      {\typeout{^^JFound AMS Symbol fonts on the system, using the
                'amssymb' package.^^J}%
       \usepackage{amssymb}%
         \let\leq=\leqslant
         \let\geq=\geqslant
      }{}
  \fi
\fi

\ifCUPmtlplainloaded \else
  \IfFileExists{amsbsy.sty}
    {\typeout{^^JFound the 'amsbsy' package on the system, using it.^^J}%
     \usepackage{amsbsy}}
    {}
\fi

\newsavebox{\astrutbox}
\sbox{\astrutbox}{\rule[-5pt]{0pt}{20pt}}

\title{Elastic deformations driven by non-uniform lubrication flows}

\author[Shimon Rubin, Arie Tulchinksy, Amir D. Gat and Moran Bercovici]
{Shimon Rubin, Arie Tulchinsky, Amir D. Gat$^1$, Moran Bercovici$^1$
 \thanks{Email address for correspondence: amirgat@tx.technion.ac.il, mberco@technion.ac.il}}

\affiliation{Faculty of Mechanical Engineering, Technion - Israel Institute of Technology, Haifa, Israel}

\date{May 23, 2016}
\begin{document}

\maketitle

\begin{abstract}
\begin{sloppypar}{The ability to create dynamic deformations of micron-sized structures is relevant to a wide variety of applications such as adaptable optics, soft robotics, and reconfigurable microfluidic devices. In this work we examine non-uniform lubrication flow as a mechanism to create complex deformation fields in an elastic plate. We consider a Kirchoff-Love elasticity model for the plate and Hele-Shaw flow in a narrow gap between the plate and a parallel rigid surface. Based on linearization of the Reynolds equation, we obtain a governing equation which relates elastic deformations to gradients in non-homogenous physical properties of the fluid (e.g. body forces, viscosity, and slip velocity). We then focus on a specific case of non-uniform Helmholtz-Smoluchowski electroosmotic slip velocity, and provide a method for determining the zeta-potential distribution necessary to generate arbitrary static and quasi-static deformations of the elastic plate. Extending the problem to time-dependent solutions, we analyze transient effects on asymptotically static solutions, and finally provide a closed form solution for a Green’s function for time periodic actuations.
}\end{sloppypar} 
\end{abstract}


\section{Introduction}

\begin{sloppypar}{Microstructures which can be dynamically deformed to desired patterns may hold promise for new applications in various fields such as adaptable optics, soft robotics, and reconfigurable microfluidics (\cite{chronis2003tunable, trivedi2008soft, unger2000monolithic}). In this work we suggest achieving such dynamic deformations by the use of non-uniform lubrication flows in a narrow gap between two plates (Hele-Shaw flow) in which at least one of the plates is elastic, and the pressure in the fluid is used to exert forces on the plate.}\end{sloppypar} 

Lubrication flows in the gap between an elastic plate and a rigid surface have been previously studied in the context of Taylor-Saffman instability  \citep{al2013two, pihler2012suppression}, viscous peeling \citep{hosoi2004peeling, lister2013viscous}, deformation due to injection of a fluid \citep{pihler2015displacement, peng2015displacement} and forced motion of an elastic plate pinned at one end \citep{trinh2014elastic}. Common to previous works is that the deformations are realized by applying pressure or force at the outer edges of the cell, and thus allow only a limited set of elastic deformations to be achieved.

\begin{sloppypar}{More complex pressure distributions can be achieved by using non-homogeneities of surface properties such as electroosmotic slip velocity, and slip length over super-hydrophobic surfaces.  For example, \cite{ajdari1995electro, ajdari1996generation} considered two dimensional electroosmotic flow (EOF) and demonstrated that undulation of a rigid plate on top of another plate having periodic distribution of zeta potential, breaks left/right symmetry, and gives rise to net flow generation between the plates. \cite{feuillebois2009effective} derived rigorous bounds on the effective slip of two-component textures in thin channels, which can guide the design of superhydrophobic surfaces for micro/nanofluidics. Recently, \cite{boyko2015flow} considered properties of non-uniform, and depth averaged EOF in a Hele-Shaw cell due to local zeta-potential distributions, and determined the necessary zeta-potential distributions to generate complex flow patterns. To the best of our knowledge, all non-uniform flow studies have been performed between rigid plates. }\end{sloppypar} 

\begin{sloppypar}{The aim of this work is to examine the utilization of pressure gradients formed due to non-uniform fluid properties in elastic Hele-Shaw configurations, as a mechanism to create desired dynamic deformations. The paper is structured as follows: In section ~\ref{section:ProbDefinition} we combine low Reynolds number and shallow geometry assumptions (lubrication approximation) with a Kirchoff-Love elastic model, to derive the governing equation for an elastic plate in a small deformation limit, driven by spatial gradients in the fluid’s properties. In section ~\ref{section:RigidWalls} we consider the limiting case of rigid walls and provide matching conditions for the pressure and stream function across piecewise constant non-homogeneity and then utilize them to obtain the resulting pressure for the case of axially symmetric non-homogeneities. We then focus on a specific case where the non-homogeneity is constituted by non-uniform EOF (i.e. non-uniform Helmholtz-Smoluchowski slip velocity) as an actuation mechanism, and in section ~\ref{section:StaticDeformation} obtain an analytical expression for the Helmholtz-Smoluchowski slip velocity distribution necessary to generate small, but otherwise arbitrary static deformations. In section ~\ref{section:NonStaticDeformation}  we determine the necessary Helmholtz-Smoluchowski velocity distribution to actuate arbitrary, asymptotically-static time-dependent elastic deformations and identify necessary conditions to obtain finite deformations. Finally, we provide a closed form solution for a Green’s function corresponding to time periodic actuations, and demonstrate its use in providing analytical and numerical solutions  for axially symmetric cases. We show that the deformation field due to time periodic oscillations of Helmholtz-Smoluchowski slip velocity within a disk, results in a deformation field of an exact dipole.  }\end{sloppypar} 

\section{Problem formulation}
\label{section:ProbDefinition}

\begin{figure}
\includegraphics[scale=0.60]{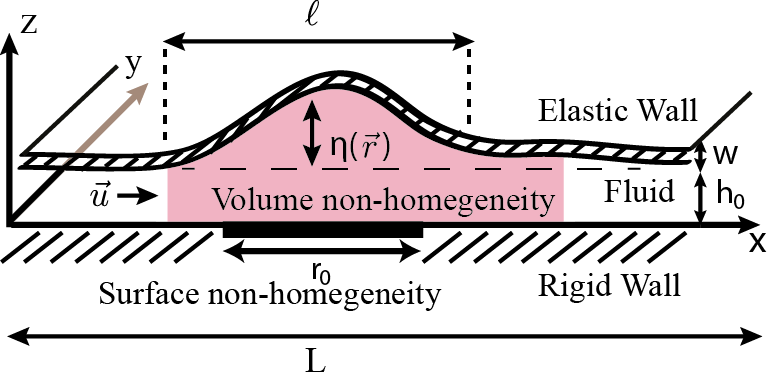}
\centering
          \caption{Schematic description of a Hele-Shaw configuration, consisting of a thin fluid layer confined between a rigid and an elastic plate of characteristic length $L$ and thickness $w$, located at an initial distance of $h_{0}$ from one another.   
The fluid is subjected to a spatial non-uniformity in its volume and/or non-uniformity of surface properties on the rigid bottom plane, on a length scale $r_{0}$, resulting in a small deformation $\eta \ll h_{0}$ (drawn not to scale) of the elastic plate, corresponding to the linearized lubrication model. The elastic plate spans a length scale $\ell$. 
The Cartesian coordinate system is chosen such that the plates are in the $x-y$ plane.}
\label{SchematicElastic}
\end{figure}

Consider a liquid layer confined between a rigid plate and an elastic plate, in a shallow geometry limit, $h_{0} \ll L$, where $h_{0}$ is the characteristic gap between the plates, and $L$ is the typical lateral in-plane dimension. 
Fig. \ref{SchematicElastic} presents a schematic illustration of the configuration and the orientation of the Cartesian coordinate system we employ, in which the plates are parallel to the $x-y$ plane, and $z$ is the normal coordinate to the plates. The total height of the elastic plate relative the rigid surface is $h=h_{0}+\eta$, where $\eta$ is the deformation of the plate.   

\begin{sloppypar}{We assume that the surface of the rigid plate or the volume between them hosts one or more regions where fluid or wall properties differ from those in the surrounding region, and that the typical size of this region is much larger than the distance between the plates, $h_{0}$, yet much smaller than the size of the plate,
\begin{equation}
	h_{0} \ll r_{0} \ll L.
\label{GeometryScales}
\end{equation}
Typical examples of surface quantities which may admit non-uniform behavior are slip velocity over surfaces with non-uniform zeta-potential or slip length over hydrophobic surfaces. Variations in bulk quantities may correspond to viscosity changes due to localized heating, or electric body forces due to application of an external electric field. 
In this section we will determine the resulting flow and pressure distribution due to the presence of such non-homogeneities.}\end{sloppypar} 

\begin{sloppypar}{ Our starting point is the Navier-Stokes equations
\begin{equation}
	\rho\left(\dfrac{\partial U_{i}}{\partial t}+ U_{j}\dfrac{\partial U_{i}}{\partial x_{j}}\right) =-\dfrac{\partial p}{\partial x_{i}}+\dfrac{\partial}{\partial x_{j}}\Bigg[ \mu \left(  \dfrac{\partial U_{i}}{\partial x_{j}}+\dfrac{\partial U_{j}}{\partial x_{i}} \right) \Bigg]+F_{i}, 
\label{NS}
\end{equation}
accompanied by the non-compressibility condition 
\begin{equation}
	\dfrac{\partial U_{i}}{\partial x_{i}}=0,
\label{continuityThreeD}
\end{equation}
where, $U_{i}$, $F_{i}$, $\rho$, $p$ and $\mu$ are the fluid velocity components, body force components, density, pressure, and dynamic viscosity, respectively. The indices $i, j$ run over the three Cartesian coordinates $x,y,z$, and summation convention over repeated indices is employed. 
We decompose all three dimensional vector fields such as fluid velocity, $\vec{U}$, and body force $\vec{F}$, into in-plane components ($x,y$) and an out-of-plane component ($z$), using the convention $\vec{U}=(\vec{u},u_{z})$ and $\vec{F}=(\vec{f},f_{z})$.
The dynamics of an elastic plate can be described by the F\"oppl von-Karman model, which accounts for both bending and tension forces in the plate. Since the ratio of the stretching to bending  free energies scales as $\eta^{2}/w^{2}$, for small deformations ($\eta \ll w$), stretching can be neglected, leading to the Kirchoff-Love model. \citep{landau1986theory} This model relates the pressure $p$ and the deformation $\eta$ according to
\begin{equation}
           \rho \dfrac{\partial^{2} \eta}{\partial t^{2}}+D \nabla^{4} \eta=p, \quad D=\dfrac{Yw^{3}}{12(1-\sigma^{2})}.
\label{LinearEquationDeformation}
\end{equation}
Here, the bending stiffness coefficient, $D$, has been expressed in terms of the plate's thickness, $w$, Young's modulus, $Y$, and Poisson's ratio, $\sigma$, and introduces an elastic time scale $\tau_{EL}$ according to 
\begin{equation}
	\tau_{EL}= \ell^{2} \sqrt{\dfrac{\rho}{D}}.
\end{equation}
We focus on shallow geometries and small Womersley and reduced Reynolds numbers, 
\begin{equation}
	\varepsilon=\dfrac{h_{0}}{r_{0}}  \ll 1; \quad \text{Wo}=\dfrac{\rho h_{0}^{2}}{\mu_{0} \tau_{f}} \ll 1; \quad \varepsilon \text{Re}=\dfrac{\rho u^{(0)}h_{0}^{2}}{\mu_{0} r_{0}} \ll 1,   
\end{equation}
where $\text{Re}=\rho  u^{(0)} h_{0}/\mu_{0}$ is the Reynolds number \citep{panton2006incompressible} and $\tau_{f}$ is the typical time-scale.
We assume that the size of the system, $L$, is much larger than the typical width of the generated deformation, $\ell$, which together with the small deformation assumption combines to }\end{sloppypar} 
\begin{equation}
	\eta \ll w \ll \ell \ll  L.
\end{equation}

Normalizing by typical scales, 
\begin{equation}
\begin{split}
	&\vec{r} \rightarrow \dfrac{\vec{r}}{r_{0}}; \hspace{0.07in} z \rightarrow  \dfrac{z}{h_{0}};  \hspace{0.07in} \vec{u} \rightarrow \dfrac{\vec{u}}{u^{(0)}};  \hspace{0.07in}
u_{z} \rightarrow  \dfrac{u_{z}}{u^{(0)}_{z}};
\hspace{0.07in}
f_{z} \rightarrow \dfrac{f_{z}}{\mu_{0} u^{(0)} r_{0}/h_{0}^{3}}
\\
         & t \rightarrow \dfrac{t}{\tau_{f}}; \quad \mu \rightarrow \dfrac{\mu}{\mu_{0}}; \quad p \rightarrow \dfrac{1}{\mu_{0} u^{(0)} r_{0}/h_{0}^{2}} p; \quad \vec{f} \rightarrow \dfrac{\vec{f}}{\mu_{0} u^{(0)} /h_{0}^{2}}; 
\end{split}
\label{scales}
\end{equation}
where $u^{(0)}$, $u_{z}$, $\mu_{0}$ are the typical magnitude of velocity along the plane, the velocity component in the normal direction and the dynamic viscosity, respectively. We can rewrite the three dimensional Navier-Stokes equations, Eq.(\ref{NS}), in non-dimensional form as
\begin{subequations}
\begin{align}
      \begin{split}
	\vec{\nabla}p-\mu\dfrac{\partial^{2} \vec{u}}{\partial z^{2}}-\vec{f}=- \text{Wo} \dfrac{\partial \vec{u}}{\partial t}-\varepsilon \text{Re} \left( \left(\vec{u} \cdot \vec{\nabla} \right)\vec{u}+u_{z}\dfrac{\partial \vec{u}}{\partial z} \right)+
    \\
  \varepsilon^{2}\dfrac{\partial}{\partial x_{\alpha}}\left( \mu \dfrac{\partial}{\partial x_{\alpha}} \vec{u} \right), \qquad \qquad \qquad \qquad \qquad
    \end{split}
\\
    \begin{split}
	\dfrac{\partial p}{\partial z}-f_{z}=-\varepsilon^{2} \text{Wo} \dfrac{\partial u_{z}}{\partial t}+\varepsilon^{2} \text{Re} \left( \left(\vec{u} \cdot \vec{\nabla} \right)u_{z}+u_{z}\dfrac{\partial u_{z}}{\partial z} \right)+
    \\
   \mu \varepsilon^{2}\dfrac{\partial^{2}u_{z}}{\partial z^{2}}+ \varepsilon^{4}\dfrac{\partial}{\partial x_{\alpha}}\left( \mu \dfrac{\partial}{\partial x_{\alpha}} u_{z} \right), \qquad \qquad \qquad
    \end{split}
\end{align}
\label{HSscaled}
\end{subequations}
where we have assumed that $\mu$ doesn't depend on the vertical coordinate $z$.
Here, the index $\alpha$ runs over the in-plane components $x,y$, and $\vec{\nabla}=( \partial / \partial x, \partial / \partial y)$ is a two dimensional gradient.
In the leading order of a small reduced Reynolds number, $\varepsilon \text{Re} \ll 1$, we can neglect the inertial terms, while in the limit of shallow geometry, $\varepsilon \ll 1$, which is equivalent to, $\partial / \partial x_{\alpha} \ll \partial/ \partial z$, we may drop the in-plane derivatives relative to the normal derivative. Together with the assumption that the time-scale $\tau_{f}$ is such that the Womersley number is small, we may overlook all inertial terms. Under these assumptions, together with an additional requirement for the normal body force component to vanish, $f_{z}=0$, the three dimensional non-linear equations, Eq.(\ref{HSscaled}), reduce to the following two dimensional vector and scalar equations 
\begin{subequations}
\begin{align}
	\vec{\nabla}p-\mu & \dfrac{\partial^{2} \vec{u}}{\partial z^{2}}-\vec{f}=0
\\
	&\dfrac{\partial p}{\partial z}=0.
\end{align}
\label{HSscaledLeading}
\end{subequations}
\begin{sloppypar}{The normal component of the velocity is assumed to satisfy the no-penetration condition on the confining walls
\begin{equation}           u_{z}\Big\vert_{z=0}=0, \quad u_{z}\Big\vert_{z=h}=\frac{\partial \eta}{\partial t}+ \vec{u} \cdot \vec{\nabla} \eta,
\label{NormalBC}
\end{equation}
while the tangential components on both planes are subject either to Helmholtz-Smoluchowski slip or Navier slip boundary conditions. Helmholtz-Smoluchowski slip occurs over electrically charged surfaces under an externally applied tangential electric field. 
The interaction of this field with the excess of net charge in the electric double layer (EDL), results in movement of fluid outside the outer edge of the EDL 
according to the Helmholtz- Smoluchowski equation \citep{hunter2001foundations}
\begin{equation}
	\vec{u} \big \vert_{(w)}=-\dfrac{\epsilon }{\mu}\zeta \vec{E},
\label{HelmSmol}
\end{equation}
where $\epsilon$, $\zeta$, $\mu$ and $\vec{E}$ are the dielectric constant of the liquid, the zeta-potential of the confining walls, 
the dynamic viscosity of the liquid, and the tangential imposed electric field, respectively. 
Note that throughout this work we assume a low Dukhin number, such that the ionic species concentrations in the bulk, as well as the associated electric field, are uniform outside the EDL. Else, non-homogeneties of zeta-potential and surface conduction, introduce non-homogeneity into the the resultant electric field in the bulk \citep{yariv2004electro, khair2008surprising}.
Slip may also emerge over hydrophobic surfaces. Such slip is typically modeled by a Navier boundary condition,
which states that the slip velocity near a flat surface, is proportional to the local velocity gradient, 
\begin{equation}
	u \bigg \vert_{(w)}=\beta \dfrac{\partial u}{\partial n} \Bigg \vert_{(w)},
\end{equation}
Here, the parameter $\beta$ is the so-called “slip length,” which for simple shear flow corresponds to the fictitious distance inside the solid where
the fluid velocity would extrapolate to zero and $n$ is the normal vector pointing into the fluid. Over the past few decades, experiments and molecular dynamics simulations have indeed confirmed that slip occurs in pressure driven flows over smooth solvophobic surfaces, with slip lengths on the order of nanometers  \citep{vinogradova1999slippage, baudry2001experimental}. 
In this work, we investigate the effect of possible surface non-homogeneities on flow and pressure fields, 
and allow both types of slip velocities to vary along the surface  }\end{sloppypar}
\begin{subequations}
\begin{align}
&\text{Helmholtz-Smoluchowski:} \hspace{0.1in} \vec{u}\Big\vert_{z=0,h}=\vec{u}_{w}^{(0,h)}(\vec{r})
\\
	&\text{Navier:} \hspace{0.1in} \vec{u}\Big\vert_{z=0}=\beta(\vec{r}) \dfrac{\partial \vec{u}}{\partial z}\Big\vert_{z=0}; \hspace{0.1 in} \vec{u}\Big\vert_{z=h}=-\beta(\vec{r}) \dfrac{\partial \vec{u}}{\partial z}\Big\vert_{z=h}.
\end{align}
\label{BC}
\end{subequations}
Integrating Eq.(\ref{HSscaledLeading}a) in the normal direction, and utilizing the boundary conditions Eq.(\ref{BC}a) and Eq.(\ref{BC}b), we obtain
\begin{equation}
	\vec{u}(\vec{r},z)=\dfrac{z(z-h)-\beta h}{2\mu} \left( \vec{\nabla}p -\vec{f} \right)+\left(1 - \dfrac{z}{h}\right)\vec{u}_{w}^{(0)}+\dfrac{z}{h}\vec{u}^{(h)}_{w}.
\label{TwoDVelocityNS}
\end{equation}
It is important to note that in Eq.(\ref{TwoDVelocityNS}) only one boundary condition can be used at a time, i.e. for Helmholtz-Smoluchowski slip velocity $\beta$ must be set to zero, while for Navier slip condition $\vec{u}_{w}^{(0,h)}$ must be set to zero. Utilizing the kinematic condition $D(z-h(\vec{r},t))/Dt=0$ (where $D$ denotes the substantial derivative), integrating Eq.(\ref{continuityThreeD}) along the normal coordinate and utilizing the no penetration boundary condition yields, 
\begin{equation}
        \dfrac{\partial h}{ \partial t} + \vec{\nabla} \cdot  ( h \langle  \vec{u} \rangle )=0 .
\label{HeleShawAverage}
\end{equation}
Here, the depth average is defined as $\langle ... \rangle = \int\limits_{0}^{h}(...)dz/\int\limits_{0}^{h}dz$.
Similarly, integrating each of the relations Eq.(\ref{TwoDVelocityNS}) and utilizing Eq.(\ref{BC}), we obtain
\begin{equation}
	\langle \vec{u} \rangle=-\left( \dfrac{h^{2}}{12\mu}+\dfrac{ \beta h}{2 \mu} \right) \left( \vec{\nabla}p -\vec{f} \right)+\langle \vec{u}_{w} \rangle,
\label{twoDEq}
\end{equation}
\begin{sloppypar}{where $\langle \vec{u}_{w} \rangle=\frac{1}{2}\left( \vec{u}^{(0)}_{w}+\vec{u}^{(h)}_{w} \right)$. Substituting the mean velocity defined in Eq.(\ref{twoDEq}) into Eq.(\ref{HeleShawAverage}), and subsequently substituting the pressure from Eq.(\ref{LinearEquationDeformation}) yields the non-linear thin film approximation
\begin{equation}
	\frac{\partial h}{\partial t}+\vec{\nabla} \cdot \Big[\left( \dfrac{h^{3}}{12\mu}+\dfrac{ \beta h^{2}}{2 \mu} \right) \left(-D  \vec{\nabla} [\nabla^{4} h] +\vec{f} \right) + h \langle \vec{u}_{w} \rangle  \Big] = 0. 
\end{equation} 
Note that in neglecting the second order time derivatives in Eq.(\ref{LinearEquationDeformation}) we have inherently assumed that relevant processes occur over time scales much longer than $\tau_{EL}$. Assuming the deformation is small, i.e. $\eta \ll h_{0}$, yields the linearized sixth-order diffusion equation 
\begin{equation}
        \dfrac{\partial \eta}{\partial t}- D h_{0}\vec{\nabla} \cdot  \left( \alpha \vec{\nabla} [\nabla^{4}\eta] \right) = - h_{0} \vec{\nabla} \cdot \left( \alpha \vec{f} + \langle\vec{u}_{w} \rangle \right).
\label{ContinuityIntegralVector3}
\end{equation}
where $\alpha=h_{0}^{2}/(12 \mu)+\beta h_{0}/(2 \mu)$.
Sixth-order diffusion equations have been employed in the study of various subjects such as magmal intrusion under a
terrestrial crust \citep{michaut2011dynamics} and interfacial instabilities under elastic membranes \citep{al2013two}.
From Eq.(\ref{ContinuityIntegralVector3}) the characteristic visco-elastic time scale, $\tau_{VE}$, is given by
\begin{equation}
	\tau_{VE}= r_{0}^{6}/\alpha D h_{0}.
\label{TauViscoElastic}
\end{equation}
We here examine configurations in which $\tau_{EL} \ll \tau_{VE}$, thus focusing our analysis on visco-elastic effects rather than propagation of elastic waves. Eq.(\ref{TauViscoElastic}) describes the time scale for the propagation of visco-elastic deformation $r_{0}$. However, if the system is actuated on a forced time scale, $\tau_{f}$, its characteristic length scale of visco-elastic dynamics is determined from Eq.(\ref{TauViscoElastic}), $r_{0}=(\tau_{f} \alpha D h_{0})^{1/6}$. In addition, if the forced time scale is smaller than $\tau_{VE}$, one must ensure that the associated $\text{Wo}$ number remains sufficiently small.}\end{sloppypar}

\section{Effect of local non-uniformities in rigid Hele-Shaw cell}
\label{section:RigidWalls}

\begin{sloppypar}{We seek to study the effect of a local variation in the fluid properties on the pressure distribution and flow field due to a combined effect of non-uniform viscosity in the bulk and non-uniform slip velocity on the surface, both described by the following axially symmetric and time independent functions,
\begin{subequations}
\begin{align}
	\mu(r)&=\mu^{(out)}+\Delta \mu \cdot  H(R-r)
\\
	\langle \vec{u}^{(w)}(r) \rangle & =\vec{u}^{(out)}+ \Delta \vec{u}^{(w)} \cdot H(R-r),
\end{align}
\label{DistributionsUMu}
\end{subequations}
which are centered at a common point, chosen as the origin of our coordinate system.
Here, $H$ stands for a Heaviside step function, $R$ is the distance from the origin at which both quantities experience a transition, and $\Delta \mu = \mu^{(in)}-\mu^{(out)}$ and $\Delta \vec{u}^{(w)}$ stand, respectively, for the difference between the values of viscosity and slip velocity in the inner $(r<R)$ and the outer $(r>R)$ regions. Assuming furthermore that only the inner region hosts a Helmholtz-Smoluchowski boundary condition, Eq.(\ref{HelmSmol}), the difference $\Delta \vec{u}^{(w)}$, between the inner and outer slip velocity, is given by 
\begin{equation}
	\Delta \vec{u}^{(w)}=-\dfrac{\epsilon \zeta^{(in)}}{\mu^{(in)}} \vec{E}.
\label{DeltaHS}
\end{equation} 
While sharp transitions of surface non-homogeneities are expected to introduce deviations from the lubrication approximation, these will be confined to a narrow spatial region of order $\varepsilon$ around the transitions \citep{brotherton2004electroosmotic}. In our analysis, we overlook these and other deviations due to sharp transition of parameters in the bulk.}\end{sloppypar}

For the case of a static elastic plate, the continuity equation Eq.(\ref{HeleShawAverage}) translates into an incompressibility condition
of the two dimensional depth averaged flow
\begin{equation}
        \vec{\nabla} \cdot  \langle \vec{u} \rangle=0.
\label{IncompresTwoD}
\end{equation}
Taking advantage of Eq.(\ref{IncompresTwoD}), we introduce a stream function $\psi$ defined via,  $\langle \vec{u} \rangle = \vec{\nabla} \times ( \psi \hat{z} )$, and derive a higher order differential equation for the pressure by applying the two dimensional divergence operator to Eq.(\ref{twoDEq}). Similarly, dividing  Eq.(\ref{twoDEq}) by $\alpha$ and applying the normal component of the curl operator eliminates the pressure. The resultant uncoupled Poisson equations for the pressure and the stream function are given by
\begin{subequations}
\begin{align}
	 \vec{\nabla} \cdot \left( \alpha \vec{\nabla} p \right) & =\vec{\nabla} \cdot \Bigg[ \langle \vec{u}_{w} \rangle + \alpha \vec{f} \Bigg]
\\
	\vec{\nabla} \cdot \left( \dfrac{1}{\alpha} \vec{\nabla} \psi \right) &= \Bigg[ \vec{\nabla}  \times \left( \dfrac{\langle \vec{u}_{w} \rangle}{\alpha} +\vec{f} \right)  \Bigg] \cdot \hat{z}.
\end{align}
\label{ProjectedEq}
\end{subequations}
Table A1 in Appendix A presents the emergent Poisson equation for several distinct cases. 

In the particular case of constant $\alpha$ and vanishing body force, the derived Eq.(\ref{ProjectedEq}) reduce to two Poisson equations which govern the case of non-uniform slip velocity, derived in \cite{boyko2015flow}.
Note that each one of the Eq.(\ref{ProjectedEq}) for pressure and stream function are formally equivalent to Gauss' law with a position dependent dielectric function, where the pressure and stream function play the role of electrostatic potential and the right hand sides of Eq.(\ref{ProjectedEq}) are interpreted as a source terms. Thus, similarly to the case of non-uniform permittivity in electrostatics \citep{landau1984electrodynamics}, the matching conditions across the common boundary for the pressure and stream function in case of a piecewise constant distribution of $\alpha$ or $\langle \vec{u_{w}} \rangle$, are derived by line integration along an infinitesimal loop across the boundary. Specifically, these matching conditions are constituted by continuity of the potential and its derivative along the tangential direction (denoted by $r_{t}$), 
\begin{subequations}
\begin{align}
	p^{(in)}=p^{(out)};& \quad  
	\psi^{(in)}=\psi^{(out)}
\\
	 \dfrac{\partial p ^{(out)}}{\partial r_{t}}=\dfrac{\partial p ^{(in)}}{\partial r_{t}};& \quad
	 \dfrac{\partial \psi ^{(out)}}{\partial r_{t}}=\dfrac{\partial \psi ^{(in)}}{\partial r_{t}},
\end{align}
\label{ContinuityPressurePsi}
\end{subequations} 
and discontinuity of its derivative along the normal direction (denoted by $r_{n}$), given by
\begin{subequations}
\begin{align}
	\alpha^{(out)}\dfrac{\partial p^{(out)}}{\partial r_{n}}-&\alpha^{(in)}\dfrac{\partial p^{(in)}}{\partial r_{n}}=
-\vec{\nabla} \cdot \Bigg[ \langle \vec{u}_{w} \rangle + \alpha \vec{f} \Bigg],
\\
\dfrac{1}{\alpha^{(out)}}\dfrac{\partial \psi^{(out)}}{\partial r_{n}}-&\dfrac{1}{\alpha^{(in)}}\dfrac{\partial \psi^{(in)}}{\partial r_{n}}=-\Bigg[ \vec{\nabla}  \times \left( \dfrac{\langle \vec{u}_{w} \rangle}{\alpha} + \vec{f} \right)  \Bigg] \cdot \hat{z}
\end{align}
\label{DielectricBCpressure}
\end{subequations}
where superscripts $(in)$ and $(out)$, correspond to the two adjacent regions.
\begin{figure*}
\includegraphics[scale=0.75]{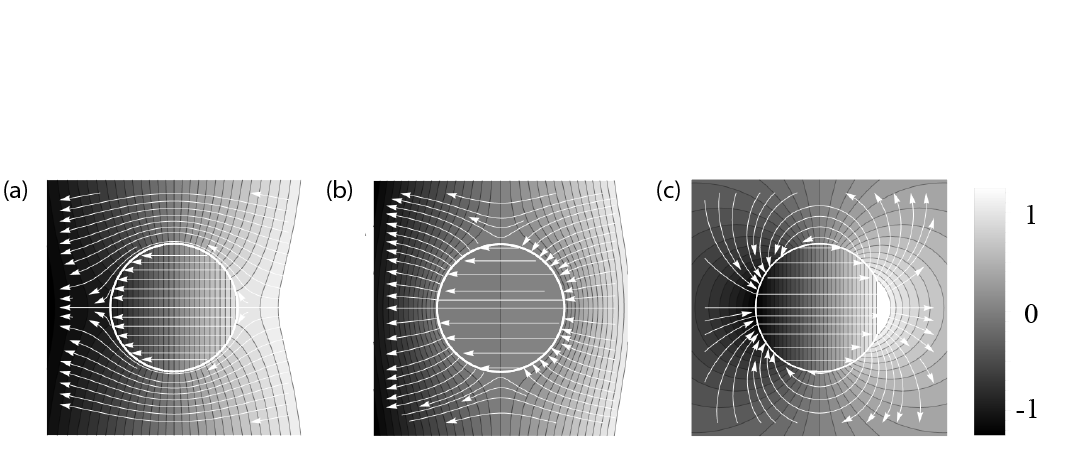}
\centering
          \caption{Analytical results showing the effect of a disk shaped non-homogeneity on the resulting pressure distribution (normalized gray-scale colormap) and flow field (white streamlines). (a,b)  The ratio of internal to external viscosity, $m$, affects the pressure distribution in the plane.  For $m>1$, the density of isobars increases within the disk and decreases outside of it. For $m<1$,  the opposite occurs and gradients are smaller within the disk and larger outside of it. In a complimentary manner, for $m>1$ streamlines are repelled from the disk region, whereas $m<1$ shows a focusing effect with streamlines at attracted to the disk. (c) A mismatch in an axial body
force, or in EOF slip velocities, between the inner and out regions of the disk results in a dipole which produces high and low pressures at the edges of the disk.}
\label{dipole}
\end{figure*}

\begin{sloppypar}{The governing equation for the pressure for the case of axially symmetric non-homogeneities in both viscosity and Helmholtz-Smoluchowski slip velocity ($\beta=0$ and $\vec{f}=0$), is explicitly written as
\begin{equation}
	\dfrac{1}{r}\dfrac{\partial}{\partial r}\left(r \dfrac{\partial p}{\partial r} \right)+\dfrac{1}{r^{2}}\dfrac{\partial^{2}p}{\partial \varphi^{2}}-\dfrac{1}{\mu} \dfrac{d \mu}{dr}\dfrac{\partial p}{\partial r}=\gamma \mu \dfrac{d }{dr} \left( \dfrac{\langle \zeta  \rangle}{\mu} \right) \cos(\varphi),
\label{GoverningAxialPressure}
\end{equation}
where $\varphi$ is the polar angle coordinate and $\gamma=12 E \epsilon/h^{2}$. The term on the right hand side suggests a dipole type solution
$p(r,\varphi)=\alpha(r)\cos(\varphi)$, which upon substitution into Eq.(\ref{GoverningAxialPressure}), together with Eq.(\ref{DistributionsUMu}) and Eq.(\ref{DeltaHS}), leads to an ordinary differential equation for $\alpha(r)$, which easily integrates and leads towards the following solution in each of the regions
\begin{equation}
	p^{(in/out)}(r,\varphi)=\left( \dfrac{a^{(in/out)}}{r}+b^{(in/out)}r \right)\cos(\varphi)
\label{ansatzpressure}
\end{equation}
where $a^{(in/out)}$, $b^{(in/out)}$ are yet to be determined coefficients. 
Regularity at the origin dictates $a^{(in)}=0$,
while $b^{(out)}=\Delta p/\Delta \ell$ can be identified as a mean pressure gradient applied along the cell, as inferred from boundary conditions at $x \rightarrow \pm \infty$. An additional relation between the coefficients, can be taken as continuity of pressure or discontinuity of its derivative along the radial direction on the curve $r=R$, constituted respectively by Eq.(\ref{ContinuityPressurePsi}a) and Eq.(\ref{DielectricBCpressure}a), leading to
the following expressions for the dipole strength, $a^{(out)}$, and pressure gradient in the inner region, $b^{(in)}$, }\end{sloppypar}
\begin{subequations}
\begin{align}
	&a^{(out)}=\dfrac{\Delta p}{\Delta \ell}R^{2}\dfrac{m-1}{m+1}+\dfrac{12 \gamma R^{2}  \zeta^{(in)}}{m+1}
\\
	&b^{(in)}=2\dfrac{\Delta p}{\Delta \ell} \dfrac{m}{m+1}+\dfrac{12 \gamma \zeta^{(in)}}{m+1},
\label{ab}
\end{align}
\end{subequations}
where $m=\mu^{(in)}/\mu^{(out)}$. 
Specific expressions for pressure and flow field can be obtained by substituting these coefficients, Eq.(\ref{ab}), into Eq.(\ref{twoDEq}) and Eq.(\ref{ansatzpressure}).
Fig.(\ref{dipole}) presents the pressure and flow fields for several cases of spatial non uniformities. Fig.(\ref{dipole}a) and Fig.(\ref{dipole}b), present the case of increasing ($m>1$)  or decreasing ($m<1$) the viscosity in the bulk, with vanishing slip velocity throughout. Under a global pressure gradient, the streamlines are repelled from the central region for $m>1$, and converge for $m<1$ (as could be anticipated due to formal analogy between position dependent $1/\mu$ in Stokes flow and position dependent dielectric constant in an electrostatic problem). In the limit $m \gg 1$ the radial component of the flow field in the outer region, which is provided in Appendix B, vanishes on the surface $r=R$, and the velocity in the inner region tends to zero.
Substituting $m=1$ into Eq.(\ref{ab}), corresponds to a case of uniform viscosity and reproduces  the electro-osmotic dipole obtained in \cite{boyko2015flow} with closed streamlines shown in Fig.(\ref{dipole}c). In Appendix C we discuss the use of the Green's function method, which allows to solve Eq.(\ref{ProjectedEq}) for more general cases.

\section{Static elastic deformations}
\label{section:StaticDeformation}

\begin{figure*}
\includegraphics[scale=0.7]{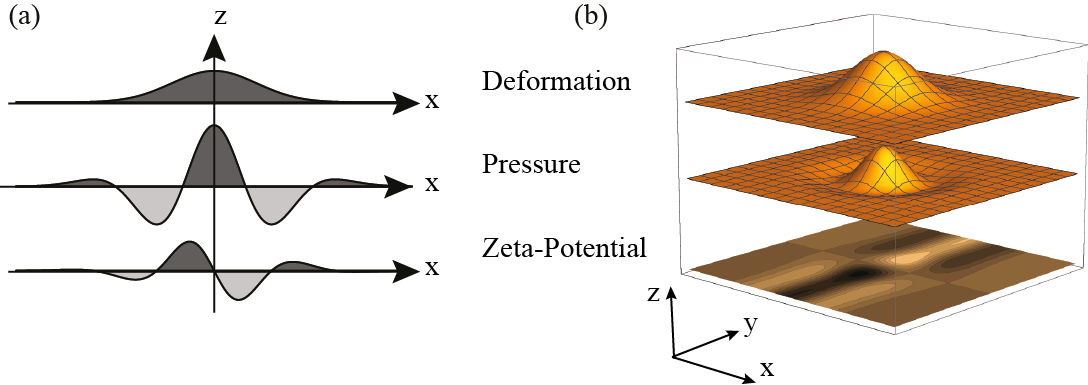}
\centering
3          \caption{Analytical results presenting the zeta-potential required for generating a desired Gaussian deformation, and the resulting pressure distribution. (a) and (b) present the 1D and 2D cases, respectively. In both cases a central region of positive pressure is bracketed by regions of negative pressure to ensure the deformation remains localized.}
\label{heigth}
\end{figure*}

\begin{sloppypar}{Consider static or quasi-static processes characterized by time scales much larger than the characteristic visco-elastic time $r_{0}^{6}/\alpha D h_{0}$, defined in Eq.(\ref{TauViscoElastic}). The corresponding governing equation for such processes then allows to neglect partial derivatives with respect to time in Eq.(\ref{ContinuityIntegralVector3}), and for the particular case of non-uniform Helmholtz-Smoluchowski slip velocity
it takes the following form
\begin{equation}
	\dfrac{D h_{0}^{2}}{12} \nabla^{6} \eta =\epsilon \vec{\nabla} \langle \zeta \rangle \cdot \vec{E}.
\label{DiffusionElasticEquationDimensionlessStatic}
\end{equation}
\begin{figure*}
\includegraphics[scale=0.7]{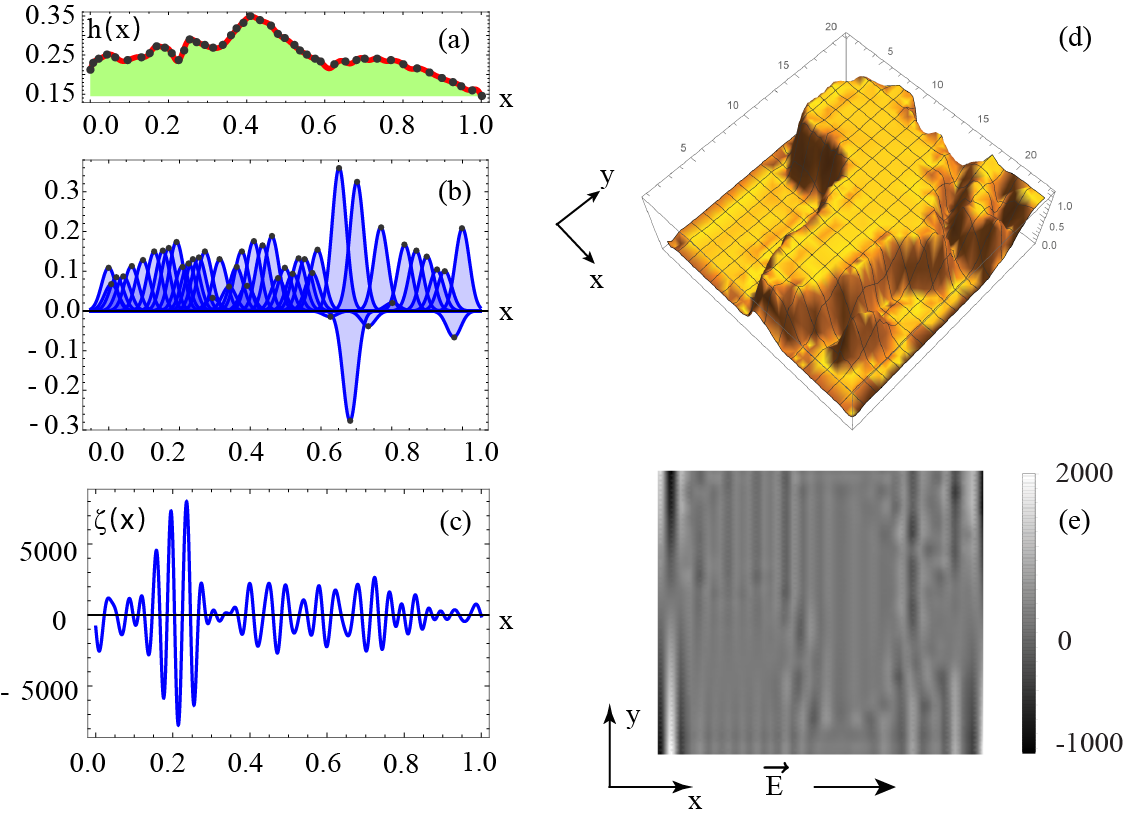}
\centering
          \caption{Analytical results demonstrating calculation of zeta-potential distributions required in order to create complex deformations. (a) presents a smooth curve of a human face profile, which is generated by superposition of $45$ Gaussians, presented in (b). (c) presents the total zeta-potential distribution required to generate this deformation. 
(d) The sum of $400$ Gaussians ($20 \times  20$ array) is used to describe a deformation corresponding to the topography of an African continent. (e) The zeta-potential distribution (colormap) in the $x-y$ plane required to achieve this deformation.}
\label{face}
\end{figure*}
Since the spatial derivative and the source terms in Eq.(\ref{ContinuityIntegralVector3}) both depend on $1/\mu$, the relation Eq.(\ref{DiffusionElasticEquationDimensionlessStatic}) between the static deformation and the corresponding zeta-potential doesn't depend on viscosity. For an arbitrary zeta potential distribution, the deformation field obtained from Eq.(\ref{DiffusionElasticEquationDimensionlessStatic})
is not necessarily localized and may create deformations with characteristic length scale of the configuration $L$.
Nevertheless, we may require a localized deformation field and calculate the corresponding zeta potential.
For an electric field $E$ which is homogeneous and oriented along the $x$ direction, we utilize Eq.(\ref{DiffusionElasticEquationDimensionlessStatic}) to express the zeta-potential as a function of the deformation $\eta$ via
\begin{equation}
	\langle \zeta(\vec{r}) \rangle =\dfrac{ Y w^{3}  h_{0}^{2}}{144  \epsilon E  (1-\sigma^{2})} \int \left( \nabla^{6} \eta \right) dx.
\label{InverseZeta}
\end{equation}
The integration introduces some function of $y$, which represents the effects of boundary conditions. In our problem we assume the boundaries are sufficiently far from the region of interest, and therefore set this function to zero. 
Hereafter we consider the setup described by Fig.(\ref{SchematicElastic}), where the non-homogeneous zeta-potential $\zeta$, resides on the bottom rigid plate. Thus, the mean zeta-potential, $\langle \zeta \rangle$, relates to the zeta-potential on the bottom plate, $\zeta$, via  $\langle \zeta \rangle=\zeta/2$. }\end{sloppypar}

Let us determine the pressure and zeta potential distribution, necessary to generate a local deformation of a Gaussian shape
\begin{equation}
          g(\vec{r},\vec{r}_{i})=\dfrac{1}{\pi \sigma^{2}}e^{- \left(\vec{r}-\vec{r}_{i} \right)^{2}/\sigma^{2}}.
\label{Gaussian}
\end{equation}
Substituting Eq.(\ref{Gaussian}) into Eq.(\ref{LinearEquationDeformation}) yields an expression for $p$, while utilizing Eq.(\ref{InverseZeta}), yields the following closed form expression for the necessary zeta-potential distribution
\begin{equation}
\begin{split}
	 &\zeta(\vec{r})= -\dfrac{8 g(\vec{r})}{\sigma^{10}} \times
\\
	&\Bigg[ y \left(12 x^{4}+4 y^{4}-26 y^{2} \sigma^{2} + 33 \sigma^{4}+6 x^{2} (2 y^{2}-9 \sigma^{2} ) \right)+
\\
	&\frac{\sqrt{\pi}}{2\sigma} e^{y^{2}/\sigma^{2}}  \text{erf}\Big[ \frac{y}{\sigma} \left(-8 x^{6}+60 \sigma^{2} x^{4} - 90 \sigma^{2} x^{2} + 15 \sigma^{6} \right) \Big]  \Bigg] .
\end{split}
\label{ZetaGaussian}
\end{equation}
Fig.(\ref{heigth}a) and  Fig.(\ref{heigth}b), respectively, present the one-, and two-dimensional zeta-potential distribution necessary to generate the corresponding Gaussian deformation.
Note that in both cases the central region with positive pressure distribution is surrounded by regions with negative pressure, which ensure that the total deformation decays to zero on a length scale $\sigma$. 

\begin{sloppypar}{To determine the zeta-potential distribution necessary to generate an arbitrary deformation, $f(\vec{r})$, we take advantage of the linearity of our problem and the finite Gaussian representation introduced at \cite{gabor1946theory}, which allows to approximate any surface as a linear combination of Gaussians centered at different points via $f(\vec{r}) \simeq \sum\limits_{i=1}^{N}c_{i}g(\vec{r};\vec{r}_{i})$,  where the index $i$  identifies a Gaussian centered at a point  $\vec{r}_{i}$. The coefficients are fixed by demanding that the sum of the Gaussian terms is equal to the values of the function $f(\vec{r})$ at some $N$ auxiliary points, leading to the following system of $N$ algebraic equations for N coefficients $c_{i}$
\begin{equation}		     f(\vec{r}_{j})=\sum\limits_{i=1}^{N}c_{i}g(\vec{r}_{j};\vec{r}_{i}); \quad \quad  i,j=1,2,...,N,
\label{SumGaussians}
\end{equation}
where for convenience we set these auxiliary points as the central points of each Gaussian. The approximation in a given domain becomes more accurate as we increase the number of Gaussian terms in the sum and decrease the distances between their central points. 
The corresponding zeta potential, $\langle \zeta(\vec{r}) \rangle$, necessary to generate a desired deformation is given by
\begin{equation}
	\langle \zeta(\vec{r}) \rangle \simeq \sum\limits_{i=1}^{N}c_{i}\langle \zeta(\vec{r};\vec{r}_{i}) \rangle,
\label{TotalZeta}
\end{equation}
where each of the $\langle \zeta(\vec{r};\vec{r}_{i}) \rangle$ terms in the sum above corresponds to the zeta-potential needed to maintain a single Gaussian deformation, given by Eq.(\ref{ZetaGaussian}). }\end{sloppypar}

For illustration, consider an example for the one- and the two-dimensional cases, where we determine the zeta-potential necessary to create desired deformations which have the shape of a human face and the African continent, respectively. Fig.(\ref{face}a) presents the shape of human face created by a sum of $45$ Gaussians, while 
Fig.(\ref{face}b) presents the set of Gaussians multiplied by the corresponding coefficients $c_{i}$.
Fig.(\ref{face}c) presents the resultant zeta-potential distribution obtained by summing the contribution needed to create each one of the Gaussians, by utilizing Eq.(\ref{TotalZeta}). Following the same steps it is possible to obtain the two-dimensional zeta-potential distribution, necessary to create arbitrary deformation profiles in 2D. Fig.(\ref{face}d) presents the result of summing $400$ Gaussians, while Fig.(\ref{face}e) presents the corresponding zeta-potential obtained by inserting the zeta-potential needed to support each one of Gaussians, Eq.(\ref{ZetaGaussian}), into the sum Eq.(\ref{TotalZeta}).

\section{Time-dependent elastic deformations}
\label{section:NonStaticDeformation}

Consider the general elastic deformations equation, Eq.(\ref{ContinuityIntegralVector3}), where the only spatial non-homogeneity
is in the Helmholtz-Smoluchowski slip velocity (i.e. $\vec{f}=0$ and $\beta=0$), and 
the source term is then the component of the zeta-potential gradient along the electric field. For convenience, we scale to dimensionless variables via $\vec{r} \rightarrow r_{0} \vec{r}$, $\eta \rightarrow \eta_{0} \eta$, $t \rightarrow \tau_{VE} t$, $E \rightarrow E_{0} E$ and $\zeta \rightarrow \zeta_{0} \zeta$.
The equation then takes the form of a sixth order diffusion equation
\begin{equation}
	\dfrac{\partial \eta}{\partial t}-\nabla^{6} \eta=\kappa \vec{\nabla} \langle \zeta \rangle \cdot \vec{E},
\label{DiffusionElasticEquationDimensionless}
\end{equation}
where $\kappa=\epsilon_{0} u_{0} \tau_{VE}/\eta_{0}$, $\epsilon_{0}=h_{0}/r_{0}$, $u_{0}=\epsilon \zeta_{0} E_{0}/\mu$ is the typical slip velocity, and $\tau_{VE}$ is defined by Eq.(\ref{TauViscoElastic}).
Similarly to Eq.(\ref{InverseZeta}), an explicit expression for the zeta-potential required to generate pre-assigned spatio-temporal elastic deformations, $\eta(\vec{r},t)$, can be obtained by expressing the corresponding zeta-potential, $\zeta(\vec{r},t)$, from the right hand side of Eq.(\ref{DiffusionElasticEquationDimensionless}), through the following integral
\begin{equation}
	\langle \zeta(\vec{r},t) \rangle=\dfrac{1}{\kappa E} \int \left( \dfrac{\partial \eta(\vec{r},t)}{\partial t} -  \nabla^{6}\eta(\vec{r},t) \right) dx,
\label{ZetaInverseElastic}
\end{equation}
where we have assumed that the external electric field is uniform and is applied along the $x$-axis. The resultant expression for $\zeta(\vec{r},t)$ in such cases is a sum of two terms. The first term expresses the zeta-potential required to sustain the rate of deformation, and a second term representing the zeta-potential of the static problem (i.e. $\partial \eta / \partial t=0$), given by Eq.(\ref{InverseZeta}).

\subsection{Asymptotically static transient deformations}

A particularly useful family of solutions are asymptotically static deformations, where $\partial \eta/ \partial t \rightarrow 0$ as $t \rightarrow \infty$.
Such deformations correspond, for example, to transitions from a non-deformed initial condition ($\eta=0$) to one of the static solutions described in the previous section. 

\begin{figure}
\includegraphics[scale=0.7]{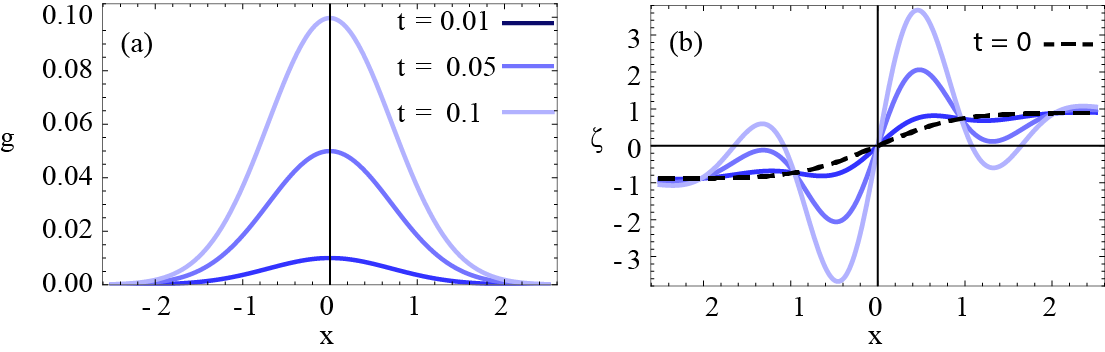}
\centering
          \caption{Analytical result showing the zeta-potential distribution required for smoothly transitioning from a non-deformed membrane to a Gaussian deformation. (a) We define the desired deformation as $\eta_{0}(\vec{r},t)=g(\vec{r})T(t)$, where $g(\vec{r})$ is a Gaussian and $T(t)$ is a smooth function which asymptotes to unity for $t \rightarrow \infty$ (here $T(t)=\tanh(t)$). (b) The required zeta potential at several early time points. 
For early times, the term associated with the rate of transition in Eq.(\ref{ZetaInverseElastic}) dominates, whereas at 
$t \rightarrow \infty$ the zeta-potential asymptotes to the static solution, as given by Eq.(\ref{InverseZeta}).}
\label{RiseOfGaussiansAndZetas}
\end{figure}

\begin{sloppypar}{For illustration, let us determine the necessary zeta-potential to actuate an asymptotically static deformation, $\eta_{0}(\vec{r},t)$, that is set in a separable form of the type $\eta_{0}(\vec{r},t)=g(\vec{r})T(t)$, where $g(\vec{r})$ is a Gaussian function and $T(t)=\tanh(t)$ (see Fig.(\ref{RiseOfGaussiansAndZetas}a)). The corresponding expression for the zeta-potential is the sum of the solution to the static problem, Eq.(\ref{InverseZeta}), multiplied by $\tanh(t)$, and of the term due to explicit dependence in time $\frac{\sqrt{\pi} \sigma}{2 \cosh^{2}(t)} \text{erf}(x/\sigma)$. As shown in Fig.(\ref{RiseOfGaussiansAndZetas}b) Note that the latter indeed tends to zero as time increases, and at late times the zeta-potential tends to the solution of the static problem. }\end{sloppypar}

\subsection{Transient deformations due to a spatially localized zeta-potential distribution}

\begin{sloppypar}{We now examine the resultant elastic deformations due to zeta-potential distributions which do not generate asymptotically static deformations. 
The governing equation for deformations in the case where the source term on the right hand side of Eq.(\ref{DiffusionElasticEquationDimensionless}) represents localized actuation is of the form 
\begin{equation}
	\dfrac{\partial \eta}{\partial t}-\nabla^{6} \eta=f(t) \delta(\vec{r}),
\label{SixthOrderDiffusionTwoDim}
\end{equation} 
where $\delta(\vec{r})$ is the Dirac delta function, and $f(t)$ is an arbitrary function of the time parameter which vanishes for $t<0$. Here and in the following, without lose of generality we set $\kappa=1$.
First, we investigate the resultant deformation $\eta_{H}(\vec{r},t)$ for the case of sudden actuation $f(t)=H(t)$, where $H(t)$ is the Heaviside function. 
Note that since in this case the source term on the right hand side of Eq.(\ref{SixthOrderDiffusionTwoDim}) doesn't contain any time scale or length scale, we expect this case to exhibit a self-similar behavior. Indeed, under scaling of time and space variables via $t \rightarrow \alpha t$ and $\vec{r} \rightarrow \beta \vec{r}$ where $\alpha$ and $\beta$ are constants, the transformation law of the Heaviside and  Dirac delta function is given by $H(\alpha t)=H(t)$ and $\delta(\beta \vec{r})=\frac{1}{\beta^{2}} \delta(\vec{r})$. Assuming furthermore that $\eta_{H}$ has no poles at $r=0$ nor at $r=\infty$,
and demanding Eq.(\ref{SixthOrderDiffusionTwoDim}) with $f(t)=H(t)$ is invariant under a similarity transformation leads to  }\end{sloppypar}
\begin{equation}
	\eta_{H}(\vec{r},t)=t^{2/3} W(\xi); \quad  \xi=\dfrac{(r/6)^{6}}{t}
\label{DimensionlessVariableDiffEq}
\end{equation}
where $W$ is a function of the dimensionless variable $\xi$ and the factor of six is introduced for convenience.
In Appendix D we derive a closed form expression for $\eta_{H}$, and discuss the Green's function solutions (i.e. solutions to Eq.(\ref{SixthOrderDiffusionTwoDim}) where $f(t)=\delta(t)$. Such solutions have also been studied in \cite{tulchinsky2015transient}).
For a fixed value of the similarity variable $\xi$, the function $\eta_{H}$ is a diverging function for increasingly large values of the variable $t$.

\subsection{Time dependent deformation due to an oscillating actuation}

We now consider time dependent deformations driven by a harmonic source with a typical 
time period $\tau_{f}=1/ \omega$. In particular we consider the dynamics after a sufficiently long time $\tau_{0}$, such that all transient solutions have decayed, i.e. $\tau_{0} > L^{2}/D h_{0} \alpha$, where $L$ is a characteristic size of the system. 
Importantly, the effect of an oscillating electric field with an angular frequency $1/\tau_{f}$, is inherently limited by the momentum diffusion time scale $\rho h_{0}^{2}/\mu$. Hence, quasi-stationarity of the flow in the alternating field regime translates into a condition,  $\omega < \mu /\rho h_{0}^{2}$, which poses an upper bound on possible angular frequency \citep{minor1997dynamic}.

\begin{figure}
\includegraphics[scale=0.72]{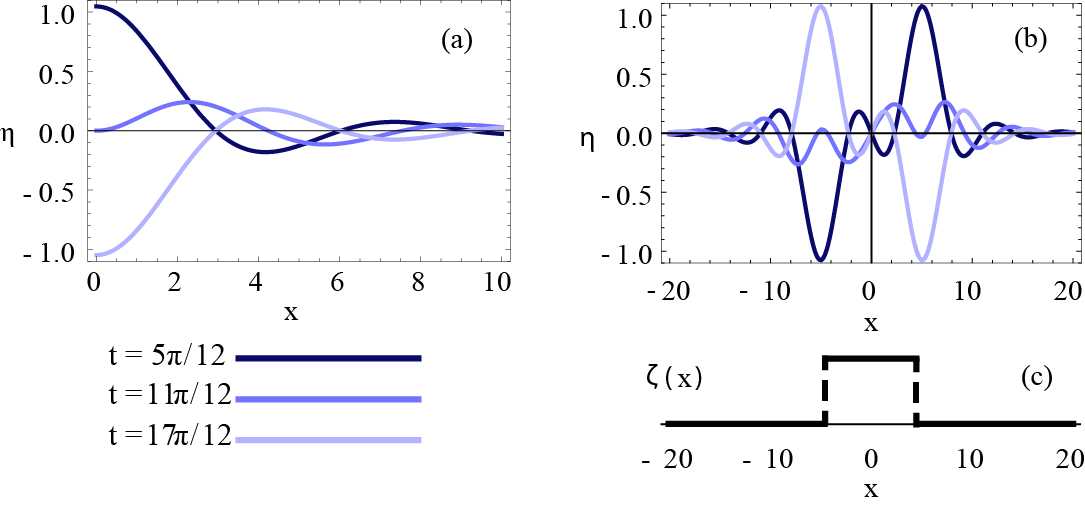} 
\centering
          \caption{ (a) Analytical results showing exponentially damped diffusion waves, due to localized oscillating source at the origin at times $t=5 \pi/12, 11 \pi/12, 17 \pi/12 $. (b) Stripe shaped distribution of zeta-potential, in which the non-zero value of zeta-potential is within a finite interval (as shown in (c)) gives rise to two opposite-sign sources in Eq.(\ref{OneDdiffusionHarmonicSource}) located at the edges of the strip. The solution is an odd function and the central point $x=0$ remains stationary throughout the motion.}
\label{PeriodicOneD}
\end{figure}

The governing Eq.(\ref{DiffusionElasticEquationDimensionless}), for the one dimensional and time dependent actuation is described by
\begin{equation}
	\dfrac{\partial \eta}{\partial t}-\dfrac{\partial^{6} \eta}{\partial x^{6}}=e^{i \omega t} \delta(x),
\label{OneDdiffusionHarmonicSource}
\end{equation}
which corresponds to a case where two adjacent half spaces, with a common boundary along the $x=0$ line, 
host different values of zeta-potential. The time dependence can stem from time varying zeta-potential or applied electric field, or both.
Assuming the solution admits harmonic time-behavior $e^{i \omega t}$, we employ the Fourier transform method, 
and derive the following representation for the deformation
\begin{equation}
	\eta(x,t)=\dfrac{e^{i \omega t}}{2 \pi}\int\limits_{-\infty}^{\infty} \dfrac{\cos(k x)}{k^{6}+ i \omega} dk.
\label{DeformationFourierRep}
\end{equation}
Utilizing the residue theorem for the choice $\omega>0$ and $x>0$, we obtain the following closed form expression for the real part of $\eta(x,t)$
\begin{equation}
\begin{split}
	\eta(x,t)= \dfrac{1}{\omega^{\frac{5}{6}}}e^{-p_{-} \omega^{\frac{1}{6}}x} \Big[  
          &  \sin \left( \omega t - p_{+} \omega^{\frac{1}{6}} x +\frac{\pi}{6} \right)+
\\
          e^{-\sqrt{3} p_{-} \omega^{\frac{1}{6}} x} & \sin \left(  \omega t - (p_{+}-p_{-}) \omega^{\frac{1}{6}}x +\frac{\pi}{6}  \right)+
\\ 
	e^{-(p_{+}-p_{-})\omega^{\frac{1}{6}}x} & \cos \left( \omega t - p_{-} \omega^{\frac{1}{6}} x \right)
\Big]
\end{split}
\label{OneDElasticGreensSolution}
\end{equation}
where $p_{\pm}$ are positive constants given by
\begin{equation}
	p_{\pm}=\dfrac{\sqrt{3} \pm 1}{2 \sqrt{2}}. 
\end{equation}
The solution Eq.(\ref{OneDElasticGreensSolution}) is a sum of three exponentially damped sinusoidal waves,
traveling with different group velocities which scale as $\omega^{\frac{5}{6}}$.
Fig.(\ref{PeriodicOneD}a) presents the solution, Eq.(\ref{OneDElasticGreensSolution}), at different times.
Asymptotic behavior at large values of $x$, singles out the first term in Eq.(\ref{OneDElasticGreensSolution}), characterized by having the smallest exponential damping factor. At the origin the motion is purely sinusoidal with angular velocity $\omega$ and normalized amplitude $\sqrt{7}$. 
Note that three different terms in the periodic solution Eq.(\ref{OneDElasticGreensSolution}) stem from the location of three simple poles in the upper complex plane. This can be compared with the solution for the simpler case of a standard diffusion equation driven by a harmonic source (i.e. Eq.(\ref{OneDdiffusionHarmonicSource}) with second order spatial derivative), which consists of a single pole in the upper complex plane and therefore a single exponentially damped term. In the limit of large $\omega$, which in the dimensionless units we use in this section corresponds to $\tau_{f} \ll \tau_{VE}$, at the origin the characteristic amplitude of the solution approaches zero as $\omega^{-\frac{5}{6}}$. Clearly, there is also a lower limit on the value of $\omega$ which is determined by the physical dimensions of the system. 

\begin{sloppypar}{As an example consider the deformation due to time periodic actuation of zeta-potential within a finite width stripe of width $2w$ (Fig.(\ref{PeriodicOneD}c)), created by a source composed of two opposite-sign Dirac-delta functions, located at the two edges of the strip.
Employing linearity of the governing equation we can then superpose the single source solution at different positions and write the resulting solution as $\eta(x+w,t)-\eta(x-w,t)$, where $\eta$ represents solution of type Eq.(\ref{OneDElasticGreensSolution}) due to each one of the sources. The solution at several time points is presented in Fig.(\ref{PeriodicOneD}b). 
Note that due to asymmetry, this solution vanishes at the origin at all times, and therefore in each one of the regions $x \leq 0$ and $x \geq 0$, the solution describes deformation due to zeta-potential distribution within a stripe of width $w$ in a presence of a fixed boundary condition (e.g. rigid wall) at $x=0$. Applying a standard images method, one can construct a solution due to an oscillating source in the presence of more complex boundaries (e.g. two walls). }\end{sloppypar}
 
\begin{figure}
\includegraphics[scale=0.6]{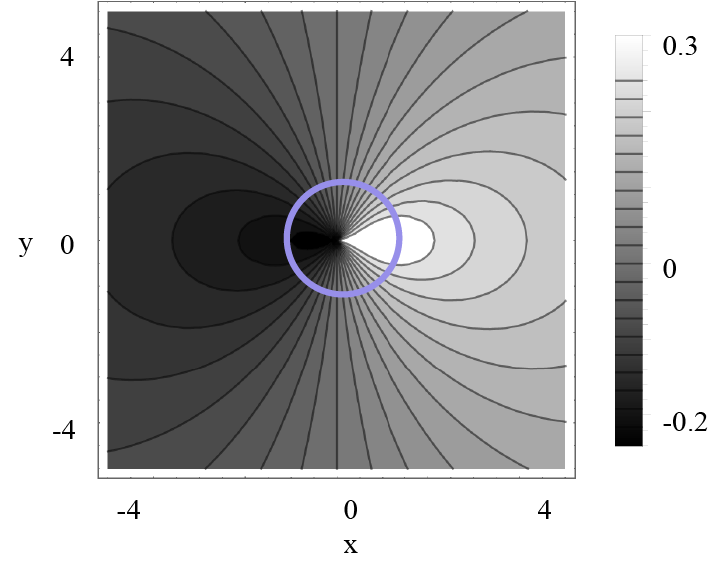}
\centering
          \caption{Numerical result showing the deformation of an elastic plate resulting from a disk-shaped zeta potential distribution of radius unity (within a blue circle) subjected to an oscillating electric field along the $x$ axis. The deformation assumes a $\cos(\varphi)$ angular dependence of a dipole, as can be anticipated from the $\cos(\varphi)$ term on the right hand side of Eq.(\ref{GoverningEqDiskPeriodic}).}
\label{PeriodicTwoD}
\end{figure}

A similar approach can be taken to extend previous results to the two-dimensional case of a spatially localized and time periodic actuation. The governing  Eq.(\ref{DiffusionElasticEquationDimensionless}) takes the following form
\begin{equation}
	\dfrac{\partial \eta}{\partial t}-\nabla^{6} \eta=e^{i \omega t} \delta(\vec{r}),
\label{PeriodicSourceSpaceLocal}
\end{equation}
and its solution can be found by standard methods of Fourier transform and complex integration, and is explicitly given by
\begin{equation}
	\eta (r,t)=e^{i \omega t} G_{0,6}^{4,0} \left(0,\frac{1}{3},\frac{2}{3},\frac{2}{3},0,\frac{1}{3} \hspace{0.025in} \Bigg \vert \hspace{0.025in} i \omega \left( \frac{r}{6} \right)^{6} \right),
\label{SolPeriodicSourceSpaceLocal}
\end{equation}
where $G^{m,n}_{p,q}$ is the so-called Meijer $G$-function \citep[see][and references within]{olver2010nist}.
\begin{sloppypar}{Note that in contrast to the one dimensional case, the source term on the right hand side of Eq.(\ref{PeriodicSourceSpaceLocal}),
doesn't correspond to a physical zeta-potential distribution. The latter follows from the fact that any two adjacent regions which host different values of zeta-potential, always introduce a zeta-potential gradient along a common boundary curve, and never along a single point.
Nevertheless, we can still utilize the solution Eq.(\ref{SolPeriodicSourceSpaceLocal}) and the superposition principle, to consider more complex study cases in 2D, such as that of a deformation due to an oscillating value of Helmholtz-Smoluchowski slip velocity within a disk of radius $r_{0}$ (either due to oscillation of zeta-potential  or of the external electrical field), subject to
\begin{equation}
	\dfrac{\partial \eta}{\partial t}-\nabla^{6}\eta=e^{i \omega t} \delta(r-r_{0}) \cos(\varphi).
\label{GoverningEqDiskPeriodic}
\end{equation}
In Appendix E, we show that the solution to Eq.(\ref{GoverningEqDiskPeriodic}) admits an angular dependence of a dipole. In particular, we show that the line integration of all $\cos(\varphi)$ weighted sources along a circle Eq.(\ref{GoverningEqDiskPeriodic}) takes the form $\eta(\vec{r},t)=e^{i \omega t} R(r) \cos(\varphi)$, and provide closed form expression for $R(r)$ in terms of Gauss hypergeometric functions.
As shown in Fig.(\ref{PeriodicTwoD}), a numerical solution can be obtained by convolving the Green's function solution Eq.(\ref{SolPeriodicSourceSpaceLocal}) with $\cos(\varphi)$ along a circle.}\end{sloppypar}

\section{Concluding remarks}

\begin{sloppypar}{In this work we studied non-uniform lubrication flows in the gap between a rigid plate and a parallel elastic plate as a mechanism to create time-varying deformation fields of the elastic plate. The approach we presented can be relevant to various applications such as adaptable optics, soft actuators, and lab-on-a-chip devices. It is instructive to estimate the resultant deformation for realistic parameters. For actuation due to EOF-driven Helmholtz-Smoluchowski slip velocity, and using typical values of a zeta potential of $ \zeta =50 [mV]$ (i.e. nearly twice a thermal voltage $V_{T} \sim 25.6 [mV]$), a plate Young modulus $Y=1 [mPa]$ corresponding to Polydimethylsiloxane (PDMS) \citep{cheng2010note}, magnitude of applied electric field $E=10^{4} [V/m]$, a gap between the plates $h_{0}=50 [\mu m]$, a plate thickness $w=10 [\mu m]$, a dielectric constant $\epsilon=80 \varepsilon_{0}$ (where $\varepsilon_{0}$ is dielectric permittivity of the vacuum), yields a deformation $\eta$ of the order of $40 [\mu m]$ distributed over a region of width $300 [\mu m]$. Naively, from the linearity of Eq.(\ref{DiffusionElasticEquationDimensionless}) one could expect that an order of magnitude increase of the zeta-potential would lead to a similar increase in the deformation. However, as can be directly obtained by the Boltzmann distribution for monovalent ions concentration in the EDL, $c_{EDL}/c_{0}=e^{\zeta/V_{T}}$, together with the fact that the ions have a finite size, there exists an upper bound for $\zeta$ above which it isn't possible to pack additional charges into the EDL. For example, a bulk solution concentration of $c_{0}=1 [mM]$, would reach the critical packing limit on the surface at approximately $\zeta=350 [mV]$, and is expected to introduce corrections even at lower values of zeta-potential \citep{bazant2009nonlinear}. Indeed, \cite{van2005field} reported saturation in EOF velocity for sufficiently high zeta-potentials.  }\end{sloppypar}

\begin{sloppypar}{In sections ~\ref{section:StaticDeformation} and ~\ref{section:NonStaticDeformation} we examined static and asymptotically static deformations. Such deformations are achieved by requiring a steady deformation field within a defined finite region and then calculating the corresponding zeta-potential distribution. In such configurations the deformations vanish outside of the deformed region, and thus boundary conditions can be applied near the deformed region without affecting the solution. Thus, the requirement $\ell \ll L$ (stating that the configuration characteristic length $L$ is much greater than the characteristic length $\ell$ of the deformed region) is unnecessary for static and asymptotically static spatially localized solutions. }\end{sloppypar}

\begin{sloppypar}{Our analysis of localized steady zeta-potential actuation yielded a solution that grows indefinitely with time, which indicates that these solutions are no longer valid after a certain time, as they fail to satisfy the underlying assumptions of our model.
In practice, there are several stabilizing mechanisms such as physical walls and the liquid's own weight, which impose restoring forces on the elastic plate and are expected to prevent the diverging behavior, though investigation of such models lays beyond the scope of this study. Within the small deformation regime used in this work, the static solutions are valid provided that $\eta_{L}/ \eta_{\ell} \sim p_{L} L^{4}⁄p_{\ell} \ell^{4} \ll 1$, where subscript $\ell$ indicates the deformation and pressure associated with the localized solution, whereas $L$ indicates quantities associated with the perturbation field over the entire configuration. }\end{sloppypar}

In this work, we focused on small deformations compared with the gap between the plates, which allows to consider one-way coupling between the fluid pressure and the elastic deformation of the plate. From our order of magnitude analysis, it appears feasible to obtain pressures which would result in deformations of few tens of microns.  
In principle, it is possible to achieve larger deformations by utilizing thinner elastic plate (at least in the Kirchoff-Love model the deformation depends on thickness in a power of three), and/or utilizing more compliant materials which may become more available with the advance of material science (e.g. materials with Young modulus $Y \sim 1 [kPa]$, reported in \cite{moraes2015supersoft}).
Modeling large deformations is an interesting extension of this work where one must to consider changes in the flow field and the electric field due to deformations, as well as to account for additional elastic forces such as the tension in the plate (which could be captured by F\"oppl-von Karman model), and also body forces such as gravity, which are expected to have an additional stabilizing effect on the dynamics of the plate. 

\section*{Acknowledgements}
This project has received funding from the European Research Council (ERC) under the European Union’s Horizon 2020 Research and Innovation Programme, grant agreement no. 678734 (MetamorphChip). We gratefully acknowledge support by the Israel Science Foundation (grant no. 818/13).  S.R. was supported in part by a Technion fellowship from the Lady Davis Foundation.


\appendix



\appendix
\section{Governing equation for several specific cases of non-homogeneous quantities}

In section 2 we derived a pair of uncoupled Poisson equations, Eq.(\ref{ProjectedEq}), for the pressure and the stream function of the depth averaged flow. The following table summarizes few cases where physical quantities experience non-uniform behavior on the surface or in the bulk, along with the corresponding governing equations for the pressure. 
\makeatletter
\let\@float@original\@float
\xpatchcmd{\@float}{\csname fps@#1\endcsname}{h!}{}{}
\makeatother

\begin{table}
\begin{tabular}{ccccc}


Quantity:&  
\begin{tabular}{@{}c@{}}  Governing Equation:   \end{tabular} & 
\begin{tabular}{@{}c@{}}  Non-uniform and vanishing quantities:  \end{tabular} & 
\begin{tabular}{@{}c@{}}     \end{tabular} 
\\ [2ex]

Helmhotlz-Smoluchowski \\ slip velocity&  
\begin{tabular}{@{}c@{}}  \quad  $\nabla^{2}p-\dfrac{12 \mu}{h^{2}} \vec{\nabla} \cdot  \langle \vec{u}_{w} \rangle =0$   \end{tabular} & 
\begin{tabular}{@{}c@{}}   $\langle \vec{u}_{w} \rangle= \langle \vec{u}_{w}  (\vec{r}) \rangle $, $\vec{f}=0$, $\beta$=0      \end{tabular} & 
\begin{tabular}{@{}c@{}}    \end{tabular} 
\\
\\ [-2.5ex]

\hline
\\
\\[-5ex] 
Viscosity&  
\begin{tabular}{@{}c@{}}   \quad  $\vec{\nabla} \cdot \left( \dfrac{1}{\mu} \vec{\nabla}p \right)=0$  \end{tabular} & 
\begin{tabular}{@{}c@{}}  $\mu=\mu(\vec{r})$,  $\vec{f}=0$, $\beta=0$, $ \langle \vec{u}_{w} \rangle =0$  \end{tabular} & 
\begin{tabular}{@{}c@{}}  \end{tabular} 
\\
\\ [-3ex]

\hline
\\
\\[-5ex] 
Navier \\ slip velocity & 
\begin{tabular}{@{}c@{}}  \quad    $\vec{\nabla} \cdot \left( \left(1+\dfrac{12 \beta}{h} \right) \vec{\nabla}p \right)=0$ \end{tabular} & 
\begin{tabular}{@{}c@{}}   $\beta=\beta(\vec{r})$, $\vec{f}=0$, $ \langle \vec{u}_{w} \rangle =0$ \end{tabular} & 
\begin{tabular}{@{}c@{}}  \end{tabular}
\\ 
\\ [-2.5ex]
\hline

\\
\\[-5ex] 
Body force& 
\begin{tabular}{@{}c@{}}  \quad   $\nabla^{2} p + \vec{\nabla} \cdot \vec{f}= 0 $ \end{tabular} & 
\begin{tabular}{@{}c@{}}    $\vec{f}=\vec{f}(\vec{r})$, $ \langle \vec{u}_{w} \rangle =0$, $\beta=0$ \end{tabular} & 
\begin{tabular}{@{}c@{}} \end{tabular}
\\ 
\\ [-2.5ex]
\end{tabular}    
\caption{Simplified governing equations for the pressure, obtained from Eq.(\ref{ProjectedEq}a) for specific choices of space dependent fluid properties.}
\label{Cases}
\end{table}

\makeatletter
\let\@float\@float@original
\makeatother

\section{Flow field for the case of non-uniform viscosity}

\begin{sloppypar}{In section 3 we derived an expression for the pressure distribution due to a disk-shaped region 
where both the viscosity and Helmholtz-Smoluchowski slip velocity take values which
are different than the values in the surrounding region (Eq.(\ref{ansatzpressure}), Eq.(\ref{ab})).
Inserting this expression into the depth-averaged momentum equation, Eq.(\ref{twoDEq}), yields the corresponding depth averaged velocity $\langle\vec{u} \rangle$. For the particular case in which only the viscosity is non-homogeneous (i.e. slip velocity remains uniform throughout the domain), the corresponding
velocity field in the outer and inner regions, respectively, is given by}\end{sloppypar}
\begin{equation}
\begin{split}
	&\langle \vec{u}^{(out)} \rangle=(u_{r}^{(out)},u_{\theta}^{(out)})=
\\
	&\dfrac{h^{2}}{12 \mu^{(out)} }\dfrac{\Delta p}{\Delta \ell}\left( \Bigg[ \dfrac{R^{2}}{r^{2}}\dfrac{m-1}{m+1}-1\Bigg] \cos(\theta),\Bigg[ \dfrac{R^{2}}{r^{2}}\dfrac{m-1}{m+1}+1\Bigg] \sin(\theta) \right),
\end{split}
\label{VelocityOuter}
\end{equation}
\begin{equation}
	\hspace{-1in} \langle \vec{u}^{(in)} \rangle=(u_{x}^{(in)},u_{y}^{(in)})=-\dfrac{h^{2}}{6 \mu^{(in)}}\dfrac{\Delta p}{\Delta \ell}\left(\dfrac{m}{m+1} ,0 \right).
\end{equation}

\section{Calculation of the pressure field by line integration of the Green's function}

In section 3 we determined the pressure due to a disk-shaped non-homogeneity of viscosity and Helmholtz-Smoluchowski slip velocity,
by solving the governing equations Eq.(\ref{ProjectedEq}) in each one of the regions, and then utilizing the matching conditions Eq.(\ref{ContinuityPressurePsi}). In a more general configurations, however, it isn't always possible to perform direct integration of Eq.(\ref{ProjectedEq}) and alternative methods are necessary.  Green's function method, which relies on the linearity of the governing equations, is particularly well suited for our problem. As was noted below Eq.(\ref{SolPeriodicSourceSpaceLocal})), physically meaningful sources in piecewise constant distributions are always distributed along curves, and the source to the Green's function itself (prior to convolution integral) doesn't have a physical meaning.

Here we demonstrate the use of the Green's function method for re-deriving the dipole solution due to a disk shaped non-homogeneity and determine the resultant pressure due to a square-shaped distribution of non-uniform Helmhotz-Smoluchowski slip velocity. To this end we utilize the following expression for the Green's function for the two dimensional Laplacian \citep{olver2010nist}
\begin{equation}
	G(\vec{r},\vec{r}^{\text{ } \prime})=\dfrac{1}{2 \pi} \ln{\vert \vec{r}-\vec{r}^{\text{ }\prime} \vert}.
\label{TwoDGreen}
\end{equation}
Consequently, the corresponding expression for the pressure $p(\vec{r})$, due to distribution of point sources along the curve $\gamma$ with a source of strength $F(\vec{r})$, is given by the following line integral along $\Gamma$
\begin{equation}
	p(\vec{r})=\gamma \int_{\Gamma} G(\vec{r}-\vec{r}^{\text{ } \prime})F(\vec{r}^{\text{ } \prime}) d\vec{r}^{\text{ } \prime},
\label{PressureGreenStatic}
\end{equation}
where $\gamma$ is a constant (see text above Eq.(\ref{ansatzpressure})). 

\subsection{Disk-shaped non-homogeneity} 
Substituting  Green's function solution, Eq.(\ref{TwoDGreen}), into Eq.(\ref{PressureGreenStatic}) and integrating along a circle of radius $R$, with $F(\vec{r}^{\prime})$ being proportional to $\cos(\varphi^{\prime})$, leads towards
\begin{equation}
	p=-\dfrac{E \zeta_{0}R}{4 \pi} \int\limits_{0}^{2 \pi} d \varphi^{\prime} \ln \left( R^{2}+r^{2}-2Rr \cos(\varphi-\varphi^{\prime}) \right) \cos \left(\varphi^{\prime} \right).
\end{equation}
Transforming the summation angle variable via, $\tilde{\varphi}=\varphi-\varphi^{\prime}$, we obtain
\begin{equation}
	p=-\dfrac{E \zeta_{0}R}{4 \pi} \int\limits_{0}^{2 \pi} d \tilde{\theta} \ln \left( R^{2}+r^{2}-2Rr \cos( \tilde{\varphi} ) \right)
\left( \cos (\tilde{\varphi}) \cos(\varphi)- \sin(\tilde{\varphi}) \sin(\varphi) \right),
\end{equation}
yielding
\begin{align}
p &=
  \begin{cases}
  \dfrac{E \zeta_{0} R^{2}}{r}\cos(\varphi)        & \text{if } R > r \\
   E \zeta_{0} r \cos(\varphi)        & \text{if }  R< r,
  \end{cases}
\end{align}
which coincides with the solution, Eq.(\ref{ab}), for the case $m=1$. 

\subsection{Square-shaped non-homogeneity}
\begin{figure}
\includegraphics[scale=0.45]{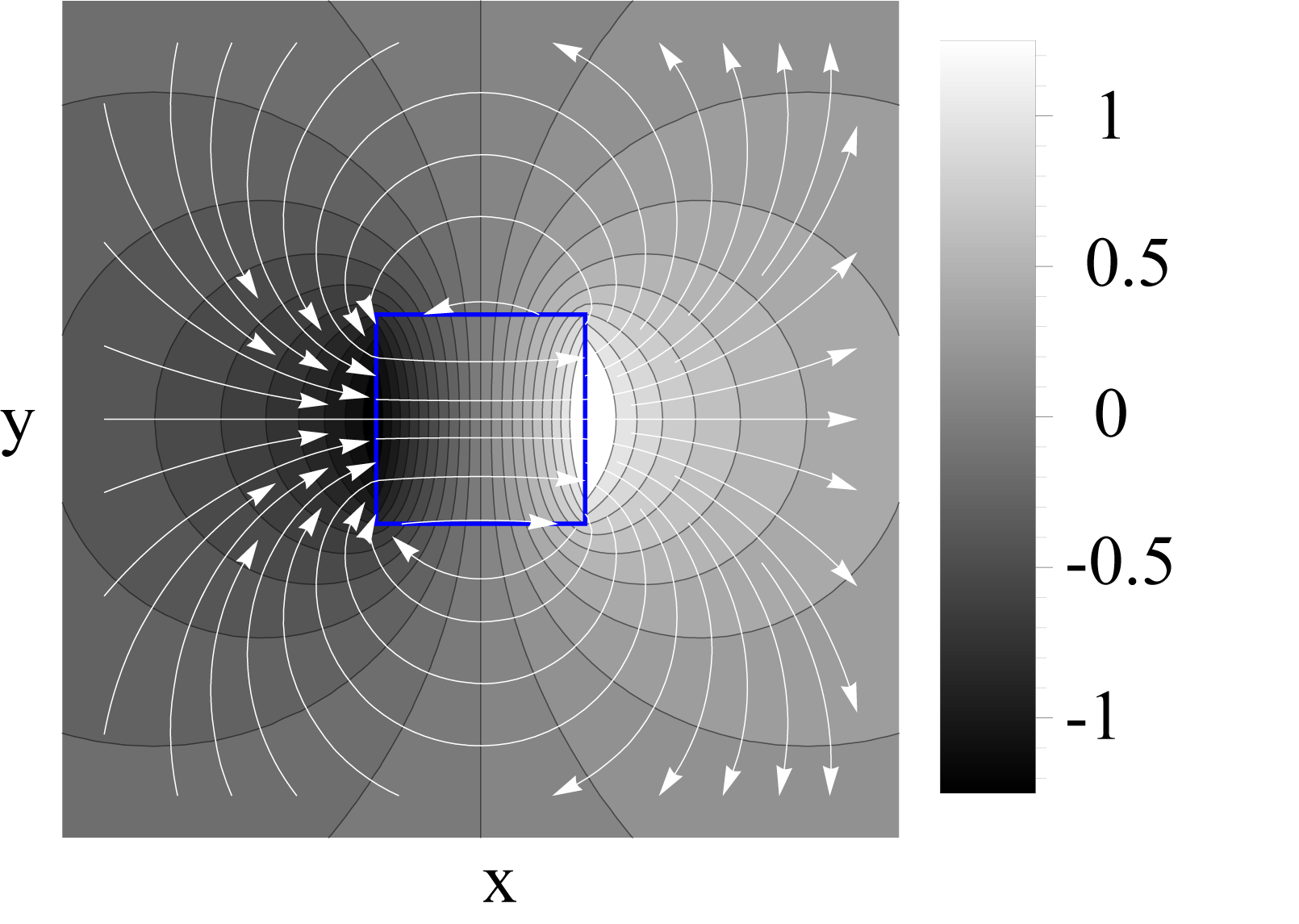}
\centering
          \caption{Analytical result showing pressure distribution (colormap) given by Eq.(\ref{PressureSquare}) and flow field (white) due to a constant zeta potential within a square region (square boundary shown in blue) under a uniform electric field along $x$ direction.}
\label{FigPressureSquare}
\end{figure}

Consider non-axially symmetric case where the Helmholtz-Smoluchowski slip velocity is zero everywhere except inside a square-shaped region. Assuming the electric field is along the $x$ direction, two sides of the square are perpendicular to the $x$ direction, and following Eq.(\ref{PressureGreenStatic}) we derive the following expression for the resultant pressure
\begin{equation}
	p(\vec{r})=f\left( \vec{r};-\frac{L}{2},\frac{L}{2}\right)-f \left( \vec{r};-\frac{L}{2},-\frac{L}{2} \right)-\left( f\left( \vec{r};\frac{L}{2},\frac{L}{2} \right)-f \left(\vec{r};\frac{L}{2},-\frac{L}{2} \right) \right)
\label{PressureSquare}
\end{equation}
where,
\begin{equation}
	f(\vec{r};\vec{r}_{0}) = \dfrac{\gamma}{2 \pi} \left((x-x_{0}) \tan^{-1} \left( \dfrac{x-x_{0}}{y-y_{0}}\right)+(y-y_{0}) \left(1-\log \left ( \sqrt{(x-x_{0})^{2}+(y-y_{0})^{2}} \right) \right) \right).
\end{equation}
Fig.(\ref{FigPressureSquare}) presents the analytical solution for the pressure distribution (colormap) and the corresponding flow field lines (white). Following similar steps the solution can be easily extended to determine the pressure and flow field due to constant values of other non-homogeneities in a polygon-shaped region. 

\section{Derivation of time-dependent Green's function and actuation due to suddenly applied source}
\label{section:TimeDependentGreen}

In section 5 we performed scaling analysis of Eq.(\ref{SixthOrderDiffusionTwoDim}) for the case of Heavisede actuation (i.e. $f(t)=H(t)$), and concluded that the corresponding solution, $\eta_{H}$, diverges in time as $t^{2/3}$. In this section we derive the specific closed form solution of the function $\eta_{H}$ (in 1D and 2D) and also numerically analyze the case of a spatially localized actuation which decays in time. 

Important for our derivation of $\eta_{H}$ is the observation that this function can be expressed as an indefinite time integral of the Green's function,  $\eta_{D}$, that solves the following equation for $t>0$
\begin{equation}
	\dfrac{\partial \eta_{D}}{\partial t} - \nabla^{6} \eta_{D} =\delta(t) \delta(\vec{r}).
\label{DiffusionDeltaGreen}
\end{equation}
Therefore, instead of directly integrating Eq.(\ref{SixthOrderDiffusionTwoDim}) (for $f(t)=H(t)$), we can integrate $\eta_{D}$, which has been derived for the 2D case in \cite{tulchinsky2015transient}. For completeness, we also provide closed form expressions for the Green's function in the 1D case.

\subsection{Transient deformation due to Heaviside actuation}

\begin{figure}
\includegraphics[scale=0.6]{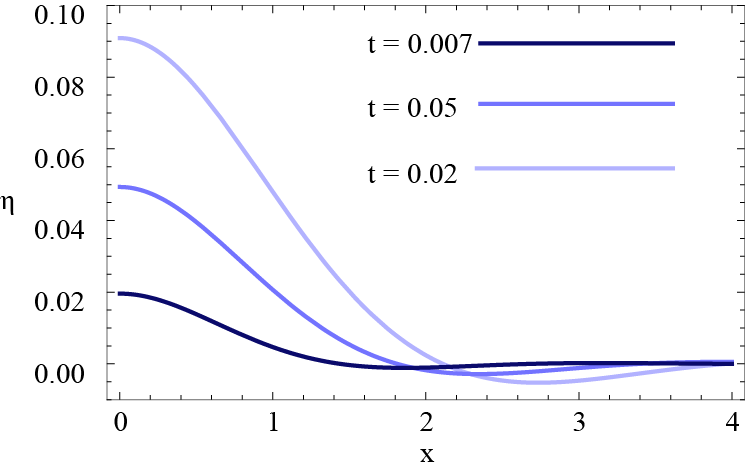}
\centering
          \caption{Analytical results presenting the 2D deformation under sudden actuation (Heaviside function), described by Eq.(\ref{HeavisideTwoDHypergeometricSolution}), at different times.}
\label{HeavisideFigure}
\end{figure}

It is useful to consider the properties of Eq.(\ref{DiffusionDeltaGreen}) under scaling transformation  $t \rightarrow \alpha t$, $\vec{r} \rightarrow \beta \vec{r}$. Similar to our discussion below Eq.(\ref{DimensionlessVariableDiffEq}), scaling, utilizing $\delta(\alpha t) \delta (\beta \vec{r}) \rightarrow \frac{1}{\alpha \beta^{2}} \delta(t) \delta(\vec{r})$ and demanding Eq.(\ref{DiffusionDeltaGreen}) is invariant under the scaling transformation yields
\begin{equation}
	\eta_{D}(\vec{r},t)=\frac{1}{t^{1/3}}U(\xi); \quad  \xi=\dfrac{(r/6)^{6}}{t}.
\label{GreenSimilarityVariable}
\end{equation}
Since $\eta_{D}$ is analytic for $t>0$, the specific form Eq.(\ref{GreenSimilarityVariable}) allows to expand it in the following series 
\begin{equation}
	\eta_{D}(\vec{r},t)=\dfrac{1}{t^{1/3}} \sum\limits_{m=0}^{\infty} c_{m} \left( \dfrac{r^{2}}{t^{1/3}} \right)^{m}.
\label{EtaDSeries}
\end{equation}
To determine closed form expression for $\eta_{D}$, we note that the Laplace transforms of Eq.(\ref{DiffusionDeltaGreen}) and the homogeneous equation
$\partial \eta_{d}/ \partial t -\nabla^{6} \eta_{d} = 0$, subject to the initial condition $\eta_{d} \vert_{t=0} = \delta(\vec{r})$, are identical, and their solutions are related via $\eta_{D}(\vec{r},t) \equiv H(t) \eta_{d}(\vec{r},t)$. The solution to the homogeneous problem, $\eta_{d}$, is easily found by summing complex separable solutions $e^{-k^{6}t} e^{i\vec{k} \cdot \vec{r}}$, for different values of the wave-vector $\vec{k}$ (identically to the 2D heat equation). Given the fact that in our problem there are no boundary conditions, we meet no restrictions on the choice of wave-vectors, and therefore  
lead towards the following integral expressions for the Green's function in 1D and 2D (for $t>0$) 
\begin{equation}
	\eta_{D}(x,t)=\dfrac{1}{2 \pi} \int\limits_{-\infty}^{\infty} e^{-k^{6}t} e^{ikx}  dk; \quad \text{in 1D},
\end{equation} 
\begin{equation}
	\eta_{D}(\vec{r},t)=\dfrac{1}{2 \pi} \int\limits_{-\infty}^{\infty} \Bigg[  \int\limits_{0}^{2 \pi} e^{-k^{6}t} e^{ikr \cos(\varphi)}d\varphi \Bigg] kdk,
	 \quad \text{in 2D},
\label{2DGreenIntegral}
\end{equation}
respectively. The integral expression for the Green's function in 2D, Eq.(\ref{2DGreenIntegral}), admits the following representation in terms of three Hypergeometric functions
\begin{equation}
\begin{split}
	\eta_{D}(\vec{r},t)=&\dfrac{\Gamma \left(\frac{1}{3} \right) }{6 t^{1/3}} { }_{0}F_{4} \left( \frac{1}{6}, \frac{1}{2},\frac{2}{3},\frac{5}{6}; -\xi  \right)
	-\frac{r^{2} \Gamma \left( \frac{2}{3} \right) }{12 t^{2/3}} { }_{0}F_{4} \left(\frac{1}{2}, \frac{5}{6},\frac{7}{6},\frac{4}{3}; -\xi  \right)-
\\ 
	&\dfrac{r^{4}}{144 t} { }_{1}F_{5} \left( 1; \frac{5}{6}, \frac{7}{6}, \frac{4}{3},\frac{3}{2},\frac{5}{3}; -\xi  \right),
\label{GreenTwoDHypergeometricSolution}
\end{split}
\end{equation}
while the coefficients, $c_{m}$, are given by \cite{tulchinsky2015transient}
\begin{equation}
	c_{m}=\dfrac{(-1)^{m}}{12 \pi \cdot {2^{2m}}}\dfrac{\Gamma \left( \dfrac{m+1}{3} \right)}{\Gamma \left( m+1 \right)^{2}}.
\end{equation}
Taking the indefinite time integral of Eq.(\ref{GreenTwoDHypergeometricSolution}) (with the position dependent integration function set to zero) we end up with the following expression for the transient solution due to Heaviside actuation
\begin{equation}
\begin{split}
	\eta_{H}(\vec{r},t)=&-\dfrac{\Gamma \left(-\frac{2}{3} \right) }{6}t^{2/3} { }_{1}F_{5} \left( -\frac{2}{3}; \frac{1}{3},\frac{1}{3},\frac{2}{3},\frac{2}{3},1; -\xi  \right)
	+\\&\frac{\Gamma \left( -\frac{1}{3} \right) }{24} t^{1/3} { }_{1}F_{5} \left(-\frac{1}{3}; \frac{2}{3},\frac{2}{3},1,\frac{4}{3},\frac{4}{3}; -\xi  \right)+
\\ 
	&\dfrac{\Gamma \left( \frac{4}{3} \right)^{2} \Gamma \left( \frac{5}{3} \right)^{2}}{384}r^{4}  G_{1,6}^{2,0}  \left( { 1 \atop 0,0,-\frac{2}{3},-\frac{2}{3},-\frac{1}{3},-\frac{1}{3} \hspace{0.025in}  }  \Bigg \vert \hspace{0.025in}  \xi  \right),
\label{HeavisideTwoDHypergeometricSolution}
\end{split}
\end{equation}
where $\xi$ is the dimensionless variable given by Eq.(\ref{DimensionlessVariableDiffEq}).
Integrating the expression Eq.(\ref{EtaDSeries}) for $\eta_{D}$ term by term (which is allowed due to uniform series convergence for $t>0$), or alternatively expanding Eq.(\ref{HeavisideTwoDHypergeometricSolution}) yields the following series representation of $\eta_{H}$
\begin{equation}
	\eta_{H}(\vec{r},t)=-t^{2/3} \sum\limits_{m=0}^{\infty} \dfrac{3c_{m}}{1+m} \left( \dfrac{r^{2}}{t^{1/3}} \right)^{m}.
\label{EtaHSeries}
\end{equation}
Fig.(\ref{HeavisideFigure}) presents $\eta_{H}$ solution, given by Eq.(\ref{HeavisideTwoDHypergeometricSolution}), at different times.

For the 1D case, the corresponding expression for 1D Green's function is given by
\begin{equation}
\begin{split}
	\eta_{D}(x,t)=&-\dfrac{\Gamma \left(\frac{7}{6} \right) }{ t^{1/6}} { }_{0}F_{4} \left( \frac{1}{3}; \frac{1}{2},\frac{2}{3},\frac{5}{6}; -\xi  \right)
	-\frac{r^{2} \sqrt{\pi/t}}{12} { }_{0}F_{4} \left(\frac{2}{3}; \frac{5}{6},\frac{7}{6},\frac{4}{3}; -\xi  \right)-
\\ 
	&\dfrac{ r^{4} \Gamma \left( -\frac{1}{6} \right) }{864 t^{5/6}} { }_{0}F_{4} \left( \frac{7}{6}; \frac{4}{3},\frac{3}{2},\frac{5}{3}; -\xi  \right),
\label{GreenOneDHypergeometricSolution}
\end{split}
\end{equation}
while the transient solution due to Heaviside actuation in 1D is given by,
\begin{equation}
	\begin{split}
&\eta_{H}(x,t)=
\\
	&-\dfrac{t^{5/6}\Gamma \left(-\frac{5}{6} \right)\Gamma \left(\frac{7}{6} \right) }{ \Gamma \left(\frac{1}{6} \right)} { }_{1}F_{5} \left( -\frac{5}{6}; \frac{1}{6},\frac{1}{3},\frac{1}{2}, \frac{2}{3}, \frac{5}{6} ; -\xi  \right)
	\\
    &- \frac{r^{4} t^{1/6} \Gamma \left(-\frac{1}{6} \right)}{144} { }_{1}F_{5} \left(-\frac{1}{6}; \frac{5}{6},\frac{7}{6},\frac{4}{3}, \frac{3}{2}, \frac{5}{3}; -\xi  \right)-
\\ 
	&\dfrac{ r^{2} t^{1/2} \sqrt{\pi} }{6} { }_{1}F_{5} \left( -\frac{1}{2}; \frac{1}{2}, \frac{2}{3}, \frac{5}{6}, \frac{7}{6},\frac{4}{3}; -\xi  \right).
\label{HeavisideOneDHypergeometricSolution}
\end{split}
\end{equation}
It is worth mentioning that since the Green's functions we found are all singular at $t=0$, it poses difficulties in constructing solutions for sources distributed along successive time moments. Specifically, direct solution of equation with a time-dependent source, $\partial \eta_{f} / \partial t - \nabla^{6} \eta_{f} = f(\vec{r},t)$, expressed through convolution of $\eta_{D}$ and $f$ (a.k.a. Duhammel's principle),
requires some regularization scheme, which isn't discussed in this work. Nevertheless, it is straightforward to create solutions by performing spatial integration of sources, acting at a given time moment.

\subsection{Transient deformation due to a time decaying source}

\begin{figure}
\includegraphics[scale=0.75]{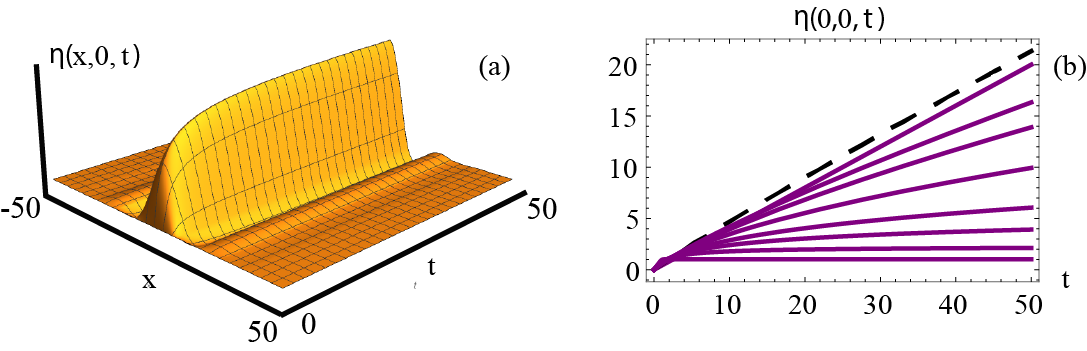}
\centering
          \caption{Numerical results showing the solution of Eq.(\ref{SixthOrderDiffusionTwoDim2}), where the parameter $b$ determines the temporal decay rate of the source and allowed to acquire a set of values. (a) shows deformation above the $x-t$ plane for $y=0$ and $b=1.5$, while (b) shows the time behavior of the origin, for the following values $b=1/10, 1/4, 1/3, 1/2, 3/4, 1, 3/2, 10$. Lower values of $b$, correspond to curves which lie closer to the dashed line $2t/5$. As $t \rightarrow \infty$ all solutions tend to zero, nevertheless prior to achieving this limit the solution may violate the initial assumption of small deformation.
}
\label{ActuationTable}
\end{figure}

Consider the resultant deformation due to spatially localized signal, multiplied by a time dependent function of the type
\begin{equation}
	\dfrac{\partial \eta}{\partial t}-\nabla^{6} \eta=\frac{e^{-r^{2}}}{1+t^{b}},
\label{SixthOrderDiffusionTwoDim2}
\end{equation} 
where $b$ is a parameter.  Fig.(\ref{ActuationTable}), presents numerical investigation of the solutions to Eq.(\ref{SixthOrderDiffusionTwoDim2}),
showing the maximal deformation at the origin as a function of time, for different values of the parameter $b$. 
The solutions show that when the actuation source doesn't decay rapidly enough, the maximal deformation grows and can eventually bring the solution out of the small deformation regime. Of course, in practice there are plenty of stabilizing mechanisms such as physical walls and liquid's own weight, which impose restoring force on the elastic plate and may prevent the diverging behavior (investigation of such models lays beyond the scope of the current study). Even without those mechanisms, for a sufficiently long time, the deformation in all cases examined decays to zero.

\section{Deformation due to time dependent actuation in a disk}

In section 5 we determined the deformation due to harmonic actuation in 1D. In this section we prove that the deformation due
to harmonic actuation in a disk must take the form of a dipole.

\begin{sloppypar}{Assume that we have solved the differential equation $L [\eta] = f(t)\delta(\vec{r})$, where $L$ is some differential operator and $f$ is some function of time argument.
Similarly to our discussion near Eq.(\ref{PressureGreenStatic}), the corresponding solution to the equation ${L [\eta_{R}] = f(t)\cos(\varphi) \delta(r-R)}$ is obtained by the following line integral along a circle of radius $R$ }\end{sloppypar}
\begin{equation}
\begin{split}
	\eta_{R}(r,\varphi)&= \int\limits_{0}^{2 \pi} d \varphi^{\prime} \eta \left( R^{2}+r^{2}-2Rr \cos(\varphi-\varphi^{\prime}) \right) \cos \left(\varphi^{\prime} \right)=
\\
	& \sum\limits_{m=0}^{\infty} c_{m} \int\limits_{0}^{2 \pi} d \tilde{\varphi}  \left( R^{2}+r^{2}-2Rr \cos( \tilde{\varphi} ) \right)^{m}
\left( \cos (\tilde{\varphi}) \cos(\varphi)- \sin(\tilde{\varphi}) \sin(\varphi) \right).
\end{split}
\label{LineIntegralDipole}
\end{equation}
In the second line of Eq.(\ref{LineIntegralDipole}) we have transformed to summation dummy variable $\tilde{\varphi}$ via $\tilde{\varphi}=\varphi-\varphi^{\prime}$, and assumed the function $\eta$ admits a series expansion $\eta(r,t)=\sum\limits_{m=0}^{\infty} c_{m}(t)r^{m}$. 
Integration term-by-term leads to
\begin{equation}
\begin{split}
	\hspace{-1in} \eta_{R}(r,\varphi) = \cos{\varphi} \sum\limits_{m=0}^{\infty} c_{m}(t) \dfrac{\sqrt{\pi }}{i r R (r^{2}-R^{2})} \dfrac{ \Gamma\left(2+m\right)}{\Gamma\left(\frac{3}{2}+m\right)} \times
\\
	\Bigg[\dfrac{r^{2}+R^{2}}{1+m} \Big[ (r-R)^{2+2m}{ }_{1}F_{2} \left( \frac{1}{2}, 1+m,\frac{3}{2}+m; -\left( \frac{r-R}{r+R}\right)^{2} \right)+
\\
	(r+R)^{2+2m} { }_{1}F_{2} \left( \frac{1}{2}, 1+m,\frac{3}{2}+m; -\left( \frac{r+R}{r-R}\right)^{2} \right) \Big] + 
\\
	\dfrac{2}{3+2m}\Big[ (r+R)^{4+2m} { }_{1}F_{2} \left( \frac{1}{2}, 2+m,\frac{5}{2}+m; -\left( \frac{r+R}{r-R}\right)^{2} \right)-
\\
	(r-R)^{4+2m} { }_{1}F_{2} \left( \frac{1}{2}, 2+m,\frac{5}{2}+m; -\left( \frac{r-R}{r+R}\right)^{2} \right) \Big] \Bigg],
\end{split}
\end{equation}  
which proves that the angular dependence of time dependent actuation in a disk must be that of a dipole and also provides a method to obtain the solution in terms of a series expansion, once the coefficients $c_{m}(t)$ are known. For a particular case of 2D harmonic actuation, the coefficients can be found by expanding the solution Eq.(\ref{SolPeriodicSourceSpaceLocal}) into a Taylor series around the origin.

\bibliographystyle{jfm}


\end{document}